\documentclass[aps,prb,showpacs,twocolumn,superscriptaddress,amsmath,amssymb,10pt]{revtex4-1}
\usepackage[T1]{fontenc}
\usepackage{graphicx,color}
\usepackage{bm}

\newcommand{\abs}[1]{\ensuremath{|#1|}}

\newcommand{\braketop}[3]{\ensuremath{\langle #1|#2|#3\rangle}}
\newcommand{\bra}[1]{\ensuremath{\langle #1|}}
\newcommand{\ket}[1]{\ensuremath{|#1\rangle}}
\newcommand{\ketbra}[2]{\ensuremath{\ket{#1}\!\bra{#2}}}

\newcommand{\sgn}{\ensuremath{\operatorname{sgn}}}
\newcommand{\up}{{\uparrow}}
\newcommand{\down}{{\downarrow}}

\newcommand{\void}[1]{}

\begin{document}
	\title{Nonlocal thermoelectricity in a Cooper-pair splitter}
	
	\author{Robert Hussein}
	\affiliation{Fachbereich Physik, Universit\"at Konstanz, D-78457 Konstanz, Germany}
	\author{Michele Governale}
	\affiliation{School of Chemical and Physical Sciences and MacDiarmid Institute for Advanced Materials and Nanotechnology, 
	Victoria University of Wellington, P.O. Box 600, Wellington 6140, New Zealand}
	\author{Sigmund Kohler}
	\affiliation{Instituto de Ciencia de Materiales de Madrid, CSIC, Cantoblanco, E-28049 Madrid, Spain}
	\author{Wolfgang Belzig}
	\affiliation{Fachbereich Physik, Universit\"at Konstanz, D-78457 Konstanz, Germany}
	\author{Francesco Giazotto}
	\affiliation{NEST, Istituto Nanoscienze-CNR, Piazza S. Silvestro 12, Pisa I-56127, Italy}
	\author{Alessandro Braggio}
	\affiliation{NEST, Istituto Nanoscienze-CNR, Piazza S. Silvestro 12, Pisa I-56127, Italy}
	\date{\today}
	
	\begin{abstract}
		We investigate the nonlocal thermoelectric transport in a Cooper-pair splitter based on a double-quantum-dot-superconductor three-terminal hybrid structure. We find 
		that the nonlocal coupling between the superconductor and the quantum dots gives rise to nonlocal thermoelectric effects which originate from the nonlocal 
		particle-hole breaking of the system.  We show that Cooper-pair splitting induces the generation of a thermo-current in the superconducting lead without any transfer of charge between the two normal metal leads. Conversely, we show that a nonlocal heat exchange between the normal leads is mediated by non-local Andreev reflection. We discuss the influence of finite Coulomb interaction and study under which conditions nonlocal power generation becomes 
		possible, and when the  Cooper-pair splitter can be employed as a cooling device.
	\end{abstract}
	
	\maketitle

	\section{Introduction}
	Hybrid superconductor devices\cite{LesovikEPJB2001a,RussoPRL2005a,Martin-RoderoAP2011a,RoddaroNanoR2011a,GiazottoNatPhys2011a,SchindelePRL2012a,RomeoNanoL2012a,DasNatureC2012a,BrauneckerPRL2013a,RossellaNatNano2014a,ChoiPRB2014a,SatoPRB2014a,DeaconNatureC2015a,MarraPRB2016a,TiiraNatureC2017a,BlasiArXiv2018a}
	are promising candidates for entanglement generation in solid-state systems and, therefore, have potential applications for  
	superconducting  spintronics,\cite{LinderNatPhys2015a} quantum information and quantum computation.\cite{MonroeNature2002a,LaddNature2010a}
	The central idea is that the electrons in a $s$-wave superconductor are in a spin-entangled state which can be made electronically accessible
	by splitting them via cross-Andreev reflection (CAR) into spatially separated  normal leads. The competing process of local Andreev 
	reflection (LAR), where the electrons tunnel into the same lead, does not directly contribute to the spatially nonlocal entanglement. 
	In order to increase the CAR fraction of the current and minimize the effect of LAR, different strategies have been adopted such as 
	employing  ferromagnetic leads,\cite{BeckmannPRL2004a,HofstetterPRL2010a,TrochaPRB2015a,WrzesniewskiJP2017a,BocianPRB2018a} 
	or including quantum dots with large intradot Coulomb repulsion.\cite{ChoiPRB2000a,RecherPRB2001a,SauretPRB2004a,
	HofstetterNature2009a,HerrmannPRL2010a,SchindelePRB2014a,FueloepPRL2015a,ProbstPRB2016a} 
	In double quantum dots with finite Coulomb repulsion, it has been discussed the possibility to induce spatially nonlocal entanglement and manipulate its symmetry by involving only the LAR process even without the nonlocal coupling.
	\cite{HusseinPRB2016a,HusseinPSSB2017a}
	
	The study of energy harvesting has also drawn much attention over the last few years.\cite{RadouskyNanotechnology2012a,RocheNatureC2015a,SothmannNanotechnology2015a,ThierschmannNatureN2015a,MastomaekiNanoR2017a} 
	Among the suggested implementations using superconductors are S-N junctions, \cite{VirtanenPRL2004a}  ferromagnet hybrid system, \cite{MachonPRL2013a,MachonNJP2014a,OzaetaPRL2014a,GiazottoPRApplied2015a,GiazottoPRL2015a,LinderPRB2016a} 
	and hybrid quantum-dot systems.\cite{WysokinskiJP2012a,HwangPRB2015a,HwangPRB2016a,TrochaPRB2017a} 
	Aspects like thermodynamic efficiencies\cite{VandenBroeckPRL2005a, MuralidharanPRB2012a,BrunnerPRE2012a,CorreaPRE2013a,MazzaNJP2014a,WhitneyPRL2014a,HoferPRB2016a,BenentiPR2017a,SeahPRE2018a} 
	and thermoelectric effects in strongly correlated quantum dots,\cite{KarwackiPRB2016a,ErdmanPRB2017a} 
	have been addressed. In particular, Machon et al.\ suggested in Ref.~\onlinecite{MachonPRL2013a} 
	that non-local thermoelectric effects in Cooper pair splitters should exist. Furthermore,
	Cao et al.\ showed in Ref.~\onlinecite{CaoAPL2015a} that Cooper-pair splitting can be achieved 
	in the absence of bias voltages by applying a thermal gradient to the normal leads.
	Inspired by this idea, we present in this work a detailed study of the nonlocal thermoelectric properties 
	of a Cooper-pair splitter taking fully into account the Coulomb interaction. Further, we discuss the possibility of nonlocal cooling 
	and power generation. Intriguingly, we show that the system still becomes a thermoelectric device  
	due to the influence of the superconducting lead, which by itself is not thermoelectrically active being intrinsically particle-hole symmetric. 
	This is essentially due to the fact that the non-local particle-hole symmetry is broken as a consequence of the thermal gradient and the three-terminal device geometry.
	
	This work is organized as follows. In Sec.~\ref{sec.:model}, we introduce our model and the formalism employed
	to calculate the thermoelectric properties. We explore the thermoelectric properties in the linear regime
	in Sec.~\ref{sec.:.linearRegime}, and compare the results to simplified effective models. Section~\ref{sec.:.powerCooling} is devoted
	to the study of nonlocal power generation and cooling. Finally, we draw our conclusions in Sec.~\ref{sec.:conclusions}.
	\begin{figure}[t]
		\includegraphics{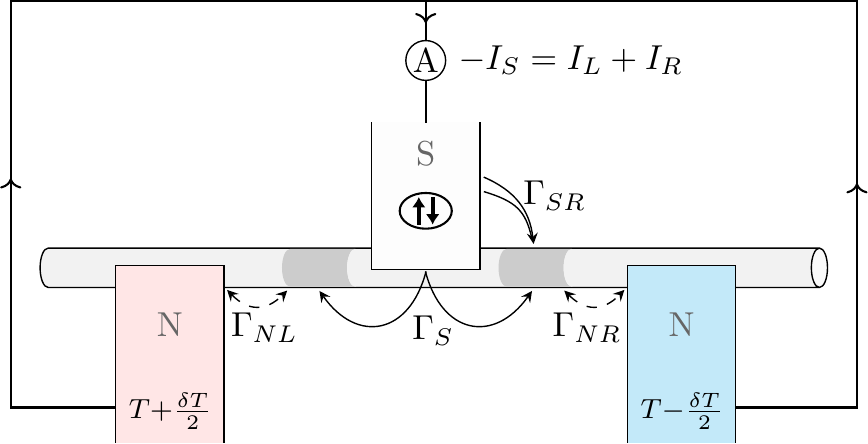}\caption{\label{fig.:setup}%
		Cooper-pair splitter circuit consisting of a double quantum dot coupled to two normal leads (N) and a superconducting one (S). 
		A temperature gradient $\delta T$ between the  normal leads induces a nonlocal current into the superconductor which may generate power for finite chemical potentials.
		}
	\end{figure}
	
	\section{\label{sec.:model}Model and master-equation}
	In this section we introduce the model of the  Cooper-pair splitter, sketched in Fig.~\ref{fig.:setup}, and the formalism employed 
	to calculate its thermoelectric properties. 
	The Cooper-pair splitter is composed of two quantum dots coupled to a $s$-wave superconductor and two normal-metal leads, 
	see Ref.~\onlinecite{HusseinPRB2016a}. 
	For a large superconducting gap, $|\Delta|\to\infty$, the subgap physics is described by the effective Hamiltonian \cite{RozhkovPRB2000a,MengPRB2009a,EldridgePRB2010a,BraggioSSC2011a,SothmannPRB2014a,HusseinPRB2016a,WeissPRB2017a,WalldorfPRL2018a}
	\begin{align}
		\begin{split}
H_{S}=H_{\textrm{DQD}} & - \sum_{\alpha=L,R} \frac{\Gamma_{S\alpha}}{2} \big( 
	d^\dagger_{\alpha\uparrow} d^\dagger_{\alpha\downarrow}+\mathrm{H.c.}
\big)\\
& -\frac{\Gamma_S}{2}\big( 
	d^\dagger_{R\uparrow} d^\dagger_{L\downarrow}- d^\dagger_{R\downarrow} d^\dagger_{L\uparrow}+ \mathrm{H.c.}
\big)  \label{eq.:Heff},
		\end{split}
	\end{align}
	where $H_{\textrm{DQD}}$ describes the double-quantum dot (DQD) system, the second term characterizes the local Cooper-pair
	tunneling between the superconductor and dot $\alpha=L,R$ with tunneling rates $\Gamma_{S\alpha}$. Here, $d^\dagger_{\alpha\sigma}$
	denotes the fermionic creation operator for an electron on dot $\alpha$ with spin $\sigma = \uparrow,\downarrow$. The last term describes 
	the nonlocal tunneling of a Cooper-pair splitting into both dots with the rate $\Gamma_S\sim\sqrt{\Gamma_{SL}\Gamma_{SR}}e^{-l/\xi}$.
	The non-local coupling $\Gamma_S$  becomes large when the distance $l$ between both quantum dots is small compared to the coherence length $\xi$ of the superconductor.
	The DQD is modeled by
	\begin{equation}
		H_{\textrm{DQD}}= %
		\sum_{\alpha,\sigma}\epsilon_{\alpha} n_{\alpha\sigma}
		+\sum_{\alpha}U_{\alpha} n_{\alpha\uparrow}n_{\alpha\downarrow}
		+U \sum_{\sigma, \sigma'} n_{L\sigma} n_{R\sigma'} \label{eq.:HDQD}
	\end{equation}
	with $\epsilon_{\alpha}$ the orbital energies and $U_\alpha$ ($U$) the intradot (interdot) Coulomb interaction;
	$n_{\alpha\sigma}=d^\dagger_{\alpha\sigma}d_{\alpha\sigma}$ is the occupation operator. 
	We note that in the limit $|\Delta|\to\infty$, the system Hamiltonian is exact in the superconducting
	coupling $\Gamma_S$.\cite{MengPRB2009a}
	This model assumes large single-level spacings in the quantum dots. Hence, a maximum of two  electrons with opposite spin can occupy each dot, and in total at most four electrons can reside in the DQD. 
	In the following, we mainly focus on the nonlocal resonance which is not substantially affected by the Coulomb interaction when $U\ll U_{\alpha}$.
	In this regime the nonlocal resonance occurs for gate voltages $\epsilon_\alpha\approx-U/2$, as discussed in Ref.~\onlinecite{HusseinPRB2016a}. Hereafter, in general we consider the case of $U=0$, but the generalization to finite interdot Coulomb 
	interaction $U\neq0$, however, is straightforward.
	
	\subsection{Master equation and transport coefficients}
	
	For the computation of particle and
	heat currents, we restrict ourselves to the sequential tunneling regime,
	$\Gamma_{NL},\Gamma_{NR}\ll k_B T_{\alpha}$, with $T_{\alpha}$ being the temperature
	of the normal lead $\alpha$. Moreover, we consider the case of weak coupling to the normal leads,
	$\Gamma_{N\alpha}\ll\Gamma_S$, and thus can express the populations $P_a$ of the
	eigenstates $\ket{a}$ of the system ($H_S\ket{a}=E_a\ket{a}$) by a Pauli-type master equation of
	the form \cite{HusseinPRB2016a} $\dot P_a = \sum_{a'} (w_{a\leftarrow
	a'}P_{a'}- w_{a'\leftarrow a}P_{a})$ with the stationary solution denoted
	as $P_a^\text{stat}$. With increasing $\Gamma_{N\alpha}$, one may have to consider off-diagonal density
	matrix elements\cite{KaiserCP2006a,DarauPRB2009a} or work in a local basis.\cite{HoferNJP2017a,GonzalezOpenSID2017a} The transition 
	rates for tunneling of an electron from the normal lead $\alpha$ to the
	respective dot ($s=+1)$ and the opposite processes ($s=-1$) are simply given by Fermi's golden rule 
	\begin{equation}
		w_{a\leftarrow a'}^{(\alpha,s)}
		=\sum_{\sigma} \Gamma_{N\alpha} 
		f_\alpha^{(-s)}(-s\omega_{aa'}) \abs{\braketop{a}{d_{\alpha\sigma}^{(-s)}}{a'}}^2 ,
		\label{eq.:w}
	\end{equation}
	with the notation $d_{\alpha\sigma}^{(-s)}$ for the electron
	creation and annihilation operators, $\omega_{aa'}=E_a -E_{a'}$, and $f_\alpha^{(s)}(\epsilon)=\{
	1+\exp[s(\epsilon-\mu_\alpha)/k_BT_\alpha] \}^{-1}$ for the Fermi function at the chemical
	potential $\mu_\alpha$.  Hereafter, we fix the chemical potential of the  superconductor to be zero, $\mu_{SC}=0$, using it as reference for the chemical potentials of the normal leads, $\mu_{\alpha}$. Hence, the total rates entering the master equation are given by 
	\begin{equation}
		w_{a\leftarrow a'} = \sum_{\alpha,s=\pm} w_{a\leftarrow a'}^{(\alpha,s)} .
	\end{equation}
	
	The electron and heat currents through the contacts correspond to the rates of 
	changes of the particle number and the energy in the corresponding lead, $I_\alpha \equiv
	e_0\langle\dot N_\alpha\rangle$ and $\dot Q_\alpha$, respectively. 
	In the sequential-tunnelling regime with the normal metal leads it is easy to write the currents in terms of the stationary populations of the DQD  $P_{a'}^{\textrm{stat}}$ and the rates $w_{a\leftarrow a'}^{(\alpha,s)}$: 
	\begin{align}
	I_{\alpha} &= \frac{e_0}{\hbar}\sum_{a,a',s=\pm}
	s w_{a\leftarrow a'}^{(\alpha,s)} P_{a'}^{\textrm{stat}}, \label{eq.:IAlpha}\\
	\dot{Q}_{\alpha} &= -\frac{1}{\hbar}
	\sum_{a,a',s=\pm} (E_a-E_{a'}) w_{a\leftarrow a'}^{(\alpha,s)}
	P_{a'}^{\textrm{stat}} -\frac{\mu_\alpha}{e_0} I_{\alpha}. \label{eq.:dotQ}
	\end{align}
	The last term in Eq.~\eqref{eq.:dotQ} reflects the fact that, in order to obtain the heat current,  one needs to subtract the net energy associated with the flux of particles at the fixed electrochemical potential $\mu_\alpha$.
	For the superconducting leads, the electric current is determined  by current conservation, that is $I_{S} = -I_{L}-I_{R}$. In the large gap limit, due to perfect Andreev heat mirroring, the heat transferred to the superconductor vanishes, i.e. $\dot Q_S=0$. This means that the heat current in the system flows only between the normal leads.
	
	In thermoelectrical systems, it is instructive to discuss the linear regime at small voltages
	and small thermal biases. Thus, the linear response of the electric currents and the heat currents 
	of our three-terminal system %
	can be described by six equations,
	which reduce to four equations, when taking into account particle and energy conservation.\cite{MazzaNJP2014a,MazzaPRB2015a,BenentiPR2017a} 
	Since for large superconducting gap heat can be exchanged only between the normal leads
	but no electric current can flow between them at the nonlocal resonance at equal chemical
	potentials, we will restrict ourself to the following relations
	\begin{align}
\label{linearcurrent}
	\delta I_S  &=  L_{11}^S \delta V + L_{12}^S \delta T ,\\
	\delta \dot  Q_R  &=  L_{21}^R \delta V + L_{22}^R \delta T,
\label{linearheat}
	\end{align}
	with four transport coefficients $L_{ij}^{S/R}$,
	$\delta V=(\mu_L+\mu_R)/2e_0$, and $\delta T=T_L-T_R$. 
	\footnote{Notice, that at the non-local resonance even for a finite voltage bias between the normal leads essentially no current
	will flow between them in the limit of strong intradot Coulomb interaction.}
	Here, $-\delta V/\delta T\big|_{I_\alpha=0}=L_{12}^S/L_{11}^S$ defines the nonlocal Seebeck coefficient, which, multiplied by the temperature difference $\delta T$ yields the 
	nonlocal Seebeck potential, and $\delta \dot Q_R/\delta T\big|_{\mu_\alpha=0}=L_{22}^R$ is the
	closed circuit thermal conductance of the right normal lead. 
	For a multi-terminal generalization, see Refs.~\onlinecite{MazzaNJP2014a,BenentiPR2017a}.
	Beyond the linear regime, Onsager coefficients of higher order\cite{SaitoPRL2007a,SaitoPRB2008a} $L_{kk'}^{S}$, $L_{kk'}^{R}$ can be calculated by recursive methods.
	\cite{HusseinPRB2014a}
	The master-equation
	formalism presented so far can be easily generalized  to compute higher-order current cumulants by using standard full-counting-statistics techniques and introducing appropriate counting variables both for the charge and energy currents.\cite{BagretsPRB2003a, BraggioPRL2006a,FlindtPRL2008a,FlindtPRB2010a,SanchezEPL2012a, GasparinettiNJP2014a,HusseinPRB2014a} Hereafter, we only consider the average currents since these are the quantities that are easily accessible experimentally.
	
	\section{\label{sec.:.linearRegime}Nonlocal thermoelectricity}
	\begin{figure*}[t]
		\includegraphics{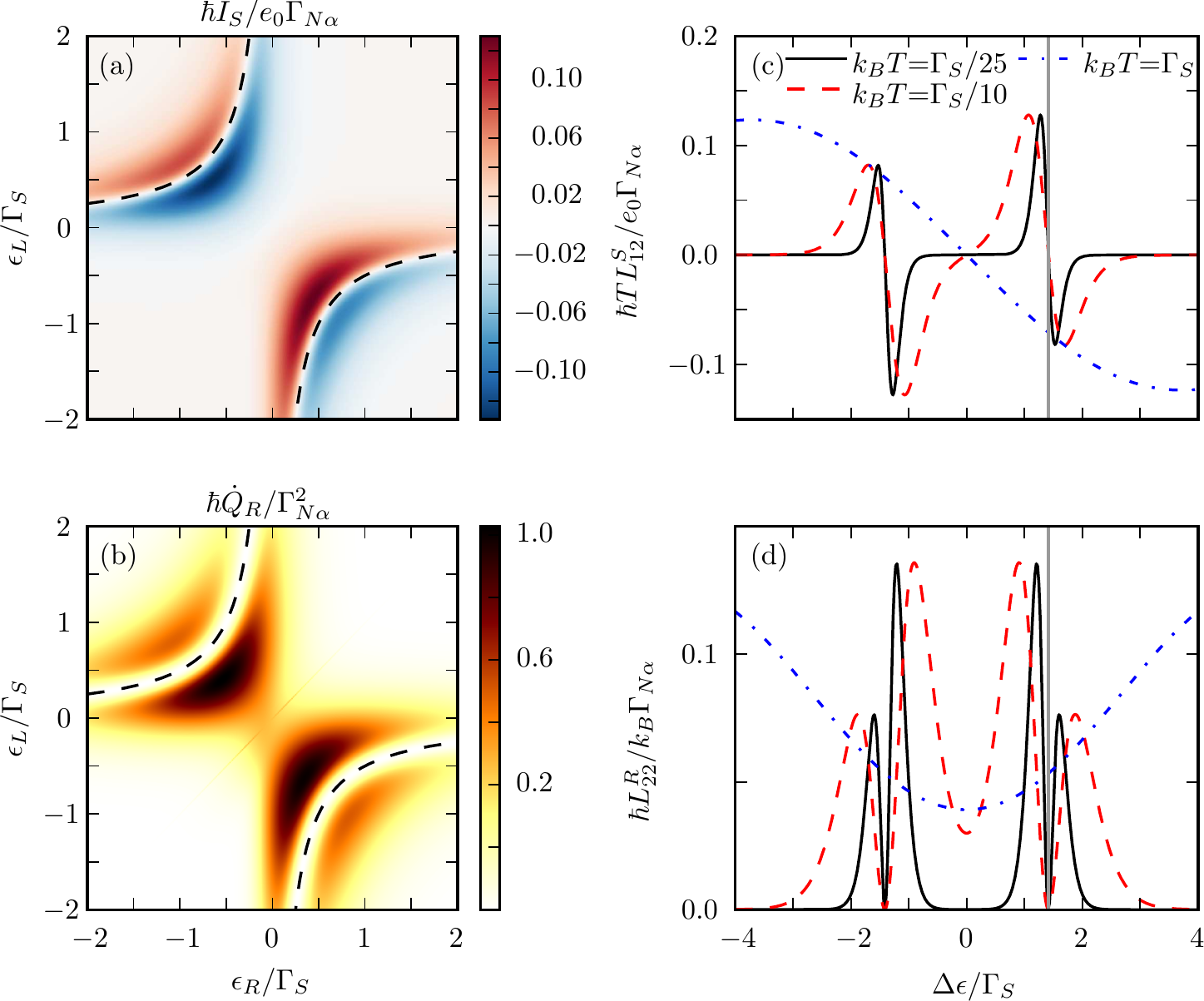}\caption{\label{fig.:nonlocalCurrent}%
		(a) Superconducting current $I_S=-I_L-I_R$ and (b) heat current ${\dot Q}_R$ through the right normal lead as a function
		of the level energies $\epsilon_L$ and $\epsilon_R$. Parameters are $k_B T_L =15 \Gamma_{N\alpha}$, $k_B T_R =5 \Gamma_{N\alpha}$,
		$\Gamma_{S}=100\Gamma_{N\alpha}$, $U=\mu_\alpha=0$, $\Gamma_{S\alpha}=\Gamma_{S}$, and  $U_\alpha\gg \Gamma_{S}$. The dashed lines indicate
		where the Andreev bound state addition energies $\big[(\epsilon_L-\epsilon_R)\pm \sqrt{2\Gamma_S^2+(\epsilon_L+\epsilon_R)^2}\big]/2$
		are resonant with the Fermi levels at $\mu=0$. 
		(c,d) Linear current coefficient $L_{12}^S$ and heat current coefficient $L_{22}^R$ in the limit $\delta T\equiv T_L-T_R\to0$ as a function
		of the detuning $\Delta\epsilon\equiv\epsilon_R-\epsilon_L$ for different average temperatures $T$. Here, the energy levels are symmetrically detuned, 
		i.e. $\epsilon_R=-\epsilon_L$. The vertical line indicates one of  the Andreev bound state addition energies.
		}
	\end{figure*}
	Hereafter, we will discuss the nonlocal thermoelectrical behavior of the Cooper-pair splitter for intradot Coulomb energies
	$U_R$ and $U_L$ much larger than any other energy scale, such that double occupancy of each indivual dot is energetically forbidden. This simplifies the system making the physics more transparent.\footnote{In principle a full investigation at finite $ U_\alpha$ can be performed, see Ref.~\onlinecite{HusseinPRB2016a}.}  
	In order to investigate thermoelectrical effects, we assume that the normal leads are at  different temperatures, $T_L>T_R$. We focus on the following non-local thermoelectrical effect: A thermal gradient between the normal leads induces
	a charge current between the superconductor and the normal leads, see Fig.~\ref{fig.:setup}, even if
	the chemical potentials of the three leads are kept equal, $\mu_{SC}=\mu_R=\mu_L=0$. 
	In the limit $U_R,U_L\gg\epsilon_R,\epsilon_L \approx 0$, the current through the superconducting lead  $I_S$ is purely induced by nonlocal Cooper-pair splitting, $I_S<0$, and recombination, $I_S>0$, respectively. In the former (latter) process Cooper pairs, consisting
	of electron singlets,  split into (recombine from) different dots. Since only non-local Andreev reflection is present, the average currents through the two normal leads are identical, $I_R=I_L=-I_S/2$, irrespectively of the lead temperatures and tunnel couplings.

	Furthermore, since we consider the situation of a large superconducting gap, $|\Delta|\to\infty$, no quasiparticle excitation can take place
	and heat transfer within the superconducting lead is forbidden. Thus, heat transfer can only occur between the two normal leads
	mediated by the superconducting lead, which operates as a perfect nonlocal Andreev mirror.
	
	In Fig.~\ref{fig.:nonlocalCurrent}(a), we show a density plot of the superconducting current $I_S$  as a function of the level energies 
	$\epsilon_R$ and $\epsilon_L$ for temperatures much smaller than the nonlocal coupling, $k_B T_\alpha\ll\Gamma_S$. We recognize 
	immediately that the current is finite for $\epsilon_R\neq\epsilon_L$.
	The current is non-vanishing close to the dashed lines corresponding  to the resonance conditions  $2\Delta E_\pm=\epsilon_L-\epsilon_R\pm\sqrt{(\epsilon_L+\epsilon_R)^2+2\Gamma_S^2}=0$. %
	
	The addition energies, $\Delta E_\pm$ correspond to processes of electron exchange at the normal leads for the model Hamiltonian
	\begin{equation}
		\label{eq:HeffReduced}
		H_{\mathrm{eff}} {=}	\sum_{\alpha\sigma}\epsilon_\alpha \Big(
		\ketbra{\alpha\sigma}{\alpha\sigma} +\frac{\ketbra{S}{S}}{2}
		\Big) -\frac{\Gamma_S}{\sqrt{2}}\big(\ketbra{0}{S}+\ketbra{S}{0}\big) \!
	\end{equation}
	following from Eq.~\eqref{eq.:Heff} when resricting it to the subspace
	involving the empty state $\ket{0}$,  the singly occupied states $\ket{\alpha\sigma}=d^\dag_{\alpha\sigma}\ket{0}$ 
	of dot $\alpha=L,R$ with spin $\sigma=\up, \down$, and the singlet state 
	$\ket{S}=\frac{1}{\sqrt{2}}\big(d^\dag_{R\up}d^\dag_{L\down} -d^\dag_{R\down}d^\dag_{L\up} \big)\ket{0}$.
	In the Hamiltonian we have omitted  the triplet states   
	as they cannot be directly coupled to the superconductor, where only singlet Cooper pairs are present. The triplet states play an important role in the high-bias regime, yielding a suppression of the current called triplet blockade.\cite{EldridgePRB2010a}
	They are also crucial in the presence of interdot tunneling in combination with spin-orbit 
	interaction.\cite{HusseinPRB2016a,HusseinPSSB2017a}
	
	In  Fig.~\ref{fig.:nonlocalCurrent}(b), we consider the heat flow from the hot to the cold normal lead. Essentially, non-vanishing heat flow occurs where the thermoelectric behavior is present, indicating that the 
	mechanism of thermo-electricity in the system is also responsible for the heat exchange. Intriguingly, here the heat exchange is mediated only by Cooper pairs, since there is no other way for an excitation to be transferred from a normal lead to another normal lead without a process involving a Cooper-pair emission or absorption at the superconducting interface.  As expected, 
	Cooper pairs cannot transfer heat to/from the superconductor, being at zero energy (ground state), but they can coherently mediate heat exchange between the normal leads. 
	We refer to this mechanism as \emph{nonlocal heat-exchange} coherently mediated by the superconducting lead. This interpretation 
	is supported by the fact that the heat exchanged is enhanced just inside the gap between the two resonances, see Fig.~\ref{fig.:nonlocalCurrent}(b). 
	Indeed, inside the gap for $\epsilon_R,\epsilon_L\approx 0$ the heat current remains finite. This is a consequence of the fact that the contributions of the two nearby resonances have opposite particle/hole character and add up. Conversely, the thermo-electrical current is suppressed in the gap since particles and holes have opposite charges and, consequently, yield opposite contributions to the thermoelectrical current. 
	It is important to stress that this heat transfer mediated by the superconductor does not affect the superconducting state and can be interpreted as a non-local version of the Andreev mirror phenomena for the heat current. 
	
	In panels (c,d) of Fig.~\ref{fig.:nonlocalCurrent}, we consider the linear regime in the temperature, $\delta T\ll T=(T_L+T_R)/2$. In this case, the linear transport coefficients depend  only on the average temperature $T$ of the leads. We  now discuss how the linear coefficients $L_{12}$ and $L_{22}$ vary with the detuning 
	$\Delta\epsilon\equiv\epsilon_R-\epsilon_L$ along the line $\epsilon_L=-\epsilon_R$ when changing the temperature but keeping fixed the nonlocal coupling $\Gamma_S$, which determines the distance between the two resonances.
	
	In Fig.~\ref{fig.:nonlocalCurrent}(c) the two central (inner) peaks
	progressively cancel each other as the temperature increases. This is a consequence of the fact that when $\Gamma_S\gtrsim k_BT$, the electron-like contribution of a resonance coexists with the 
	hole-like contribution of the other resonance. %
	This competition reduces the total thermoelectric current.
	Once $k_BT=\Gamma_S$ the two resonances merge and behave as a single resonance. Panel (d) shows the corresponding linearized heat current coefficient $L_{22}^R$. For well separated peaks, $\Gamma_S\gg k_BT$, the
	heat current essentially vanishes at the resonances, where also the current coefficient vanishes. In the situation when the peaks are in proximity,  
	$\Gamma_S\gtrsim k_BT$, they add up constructively at $\Delta\epsilon\approx 0$.
	When $k_BT=\Gamma_S$ again the thermal behavior resembles the contribution of a single QD resonance. 

	\begin{figure}[t]
		\includegraphics{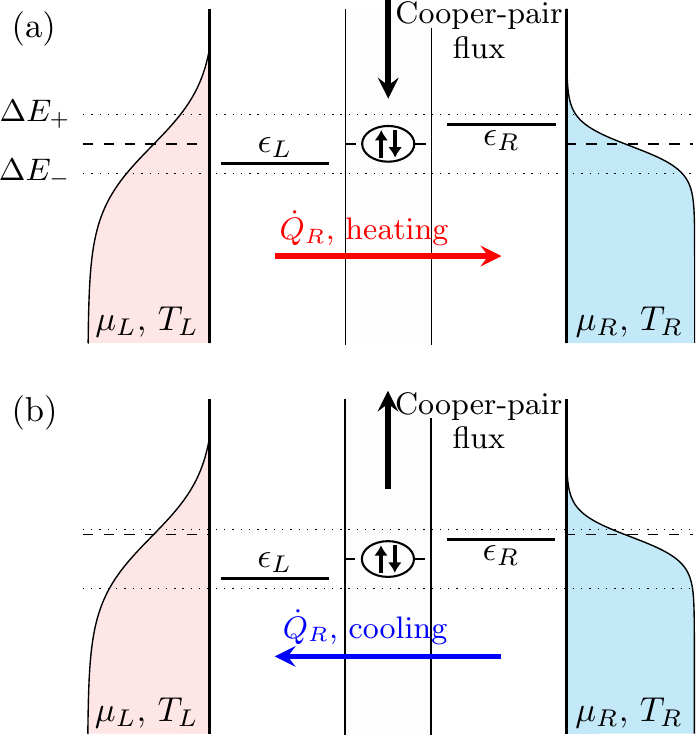}		\caption{\label{fig.:sketchHeatingCooling}%
		(a) Cooper-pair  splitter configuration for zero chemical potentials, $\mu_L=\mu_R=0$, leading to a heating of the right normal lead
		(red arrow) and a net flux of Cooper-pairs from the superconductor into the normal leads (black arrow). (b) Configuration for finite chemical potentials, $\mu_L=\mu_R>0$. A net flux of Cooper-pairs flows into the superconductor (black arrow) against the intrinsic thermo-current leading to a cooling of the right normal lead (blue arrow).
		}
	\end{figure}
	\begin{figure}[t]
		\includegraphics{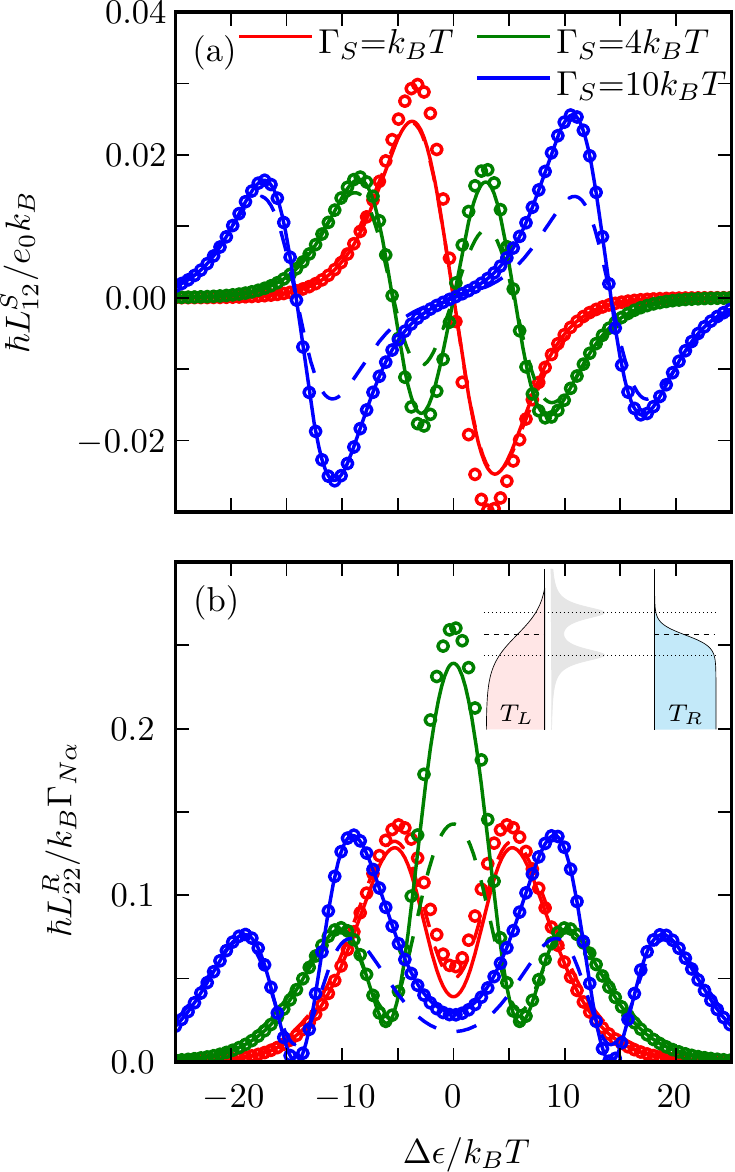}\caption{\label{fig.:linearRegime}%
		(a) Linear current and (b) heat current coefficients  $L_{12}^S$ and  $L_{22}^R$ as a function
		of the detuning $\Delta\epsilon$ with the average temperature $T=5\Gamma_{N\alpha}/k_B$ and all other parameters
		as in Fig.~\ref{fig.:nonlocalCurrent}(a). The solid lines depict the transport coefficients obtained
		from the full master equation, the circles correspond to the reduced master equation, Eqs.~\eqref{eq:LinCoeffsVSym_LS12} and \eqref{eq:LinCoeffsVSym_LR22},
		and the dashed lines correspond to the mapped model, see Eqs.~\eqref{eq.:effLinCoeffs}. The latter transport coefficients 
		are scaled by a factor of $\alpha=0.37$, such, that the effective linear current coefficient agrees with the one of the full model for $\Gamma_S=k_BT$. The inset sketches this mapped model of a single quantum dot with its onsite energies at the addition energies $\Delta E_\pm$ of the Cooper-pair splitter in the CAR regime. 
		}
	\end{figure}
	\begin{figure*}[t]
		\includegraphics{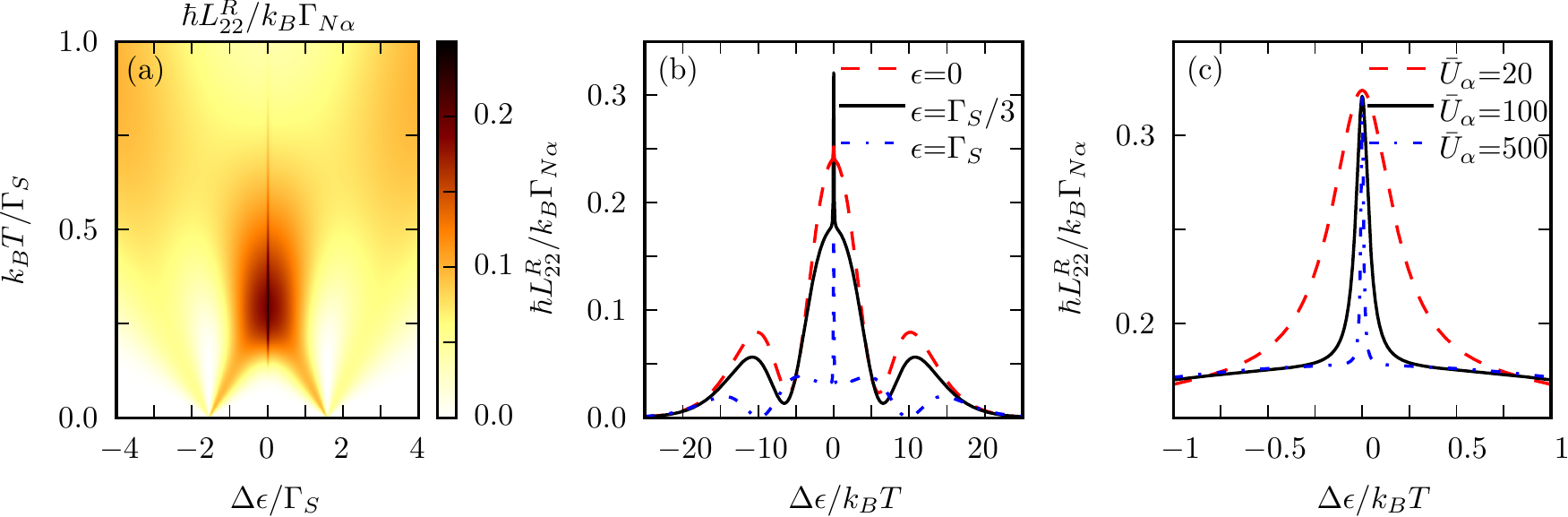}		\caption{\label{fig.:finiteUEffect}%
		(a) Linear heat current coefficient $L_{22}^R$ (which for $\delta V=0$ is proportional to the heat current $\dot Q_R$) 
		as a function of the detuning $\Delta\epsilon$ and the average temperature $T$
		for finite intradot Coulomb energy $U_\alpha=100\Gamma_{S\alpha}$, $\Gamma_{S}=\Gamma_{S\alpha}/3=4k_BT$ and $k_BT=10\Gamma_{N\alpha}$.
		Here, the level energies $\epsilon_L=\epsilon -\Delta\epsilon/2$ and $\epsilon_R=\epsilon +\Delta\epsilon/2$ are centered around the
		average value $\epsilon=\Gamma_S/3$. (b)  Dependence on the average level energies $\epsilon$ and (c) the interdot couplings 
		${\bar U}_\alpha\equiv U_\alpha/\Gamma_{S\alpha}$. 
		}
	\end{figure*}
	Let us develop a physical picture to explain the behavior of the linear thermoelectric coefficients in the limit of large intradot Coulomb interaction, for equal chemical potentials, $\mu\equiv\mu_L=\mu_R$, and in the presence of a temperature gradient 
	between the two normal leads, $T_L>T_R$.  Figure \ref{fig.:sketchHeatingCooling}(a) depicts the level structure of the double quantum-dot system in the situation 
	where the chemical potentials of the normal leads coincide with the one of the superconductor. 
	For $\epsilon_L,\epsilon_R\approx0$ the singlet state 
	mixes with the empty state forming non-local Andreev bound states shared between the two dots due to nonlocal Cooper-pair tunneling with the central superconducting lead.\footnote{The simplicity of  this picture is due to the absence of the double occupancy of the individual dots, which is energetically forbidden by the strong interaction.} 
	The electron tunneling with the  normal leads determines transition between the DQD states. 
	The dotted lines indicate the
	Andreev bound state addition energies $\Delta E_\pm$, while the solid lines indicate the orbital 
	energy levels of the quantum dots. 
	
	For the case under consideration ($T_L>T_R$), when  more electrons are above the Fermi 
	level of the left normal lead than holes below the Fermi level of the right normal lead, electrons tunnel via the Andreev bound-state channel 
	into the superconductor and form Cooper pairs. At the same time Cooper pairs can split in an opposite process and tunnel into the normal leads.  
	The difference of both processes yields a net thermoelectrical current when the normal leads have different temperatures. 
	This effect is a direct consequence of the nonlocal particle-hole asymmetry induced by the structure.

	\subsection{Linear regime and effective models}
	In Fig.~\ref{fig.:linearRegime}, we show in solid lines the dependence of the  linear coefficients on the non-local coupling $\Gamma_{\text{S}}$ keeping fixed the temperature $T$. This behavior can be roughly interpreted by mapping the Cooper-pair splitter in the CAR regime to a simplified model of a single quantum dot
	with two resonances located at the addition energies of the Andreev bound states,
	see the inset of Fig.~\ref{fig.:linearRegime}. We observe that the thermoelectric behaviour of this simple model 
	resembles the thermoelectric behaviour of the full one. Its main aim 
	is to show, that the linear behaviour of the full system is not too different from
	the conventional thermoelectrical properties of a quantum dot system;
	even though the former is mainly characterized by the discussed  nonlocality in the heat and charge transport which cannot be present in a simple quantum dot.
	The thermoelectrical coefficients for the simplified model can be expressed in the Landauer-B\"uttiker formalism as\cite{BeenakkerPRB1991a,TurekPRB2002a,NakpathomkunPRB2010a,DubiRMP2011a,EltschkaPSSA2016a} 
	\begin{align}
\label{eq.:effLinCoeffs}
	L_{k+1,2}^{\textrm{eff}}
	 &\equiv\frac{(2e_0)^{1-k}}{h} \int_{-\infty}^\infty dE\, \frac{ E^{k+1}\tau(E) }{4 k_BT^2\cosh^2\big(E/2k_BT\big)}. 
	\end{align}
	The additional factor $2$ for the electron current ($k=0$) takes into account that in an Andreev process the current is doubled ($I_S=-2I_R$). The transmission function is modeled by two Lorentzians located at the Andreev bound state energies
	\begin{align}
	 \tau(E) \propto
	 \sum_{s=\pm}\frac{\gamma}{
	 	(E-\Delta E_s)^2+ (\gamma/2)^2
	 }.
	\end{align}
	Finally we arrive at $L_{k+1,2}^{\textrm{eff}}=\frac{\alpha}{\hbar} (2e_0)^{1-k}\sum_{s=\pm}A_k(\Delta E_s)$ with $\alpha$ an overall scaling factor and for $k=0,\pm1$ the function
	\begin{equation}
		A_k(\Omega)= \sum_{s=\pm} \frac{\omega_s^{k+1}(\Omega)}{k_B (2\pi T)^2}\Psi'\Big(
		\frac{1}{2} - \frac{si\omega_s(\Omega)}{2\pi k_BT} 
		\Big)+\frac{\gamma\delta_{k,1}}{2\pi T},
	\end{equation}
	which collects the contributions from the poles of the Fermi function and the poles $\omega_s(\Omega)=\Omega+si\gamma/2$
	of the Lorentzians. Here, $\Psi$ denotes the digamma function. %
	
	The reader should be aware that in the quantum dot model the thermo-electrical current and the thermal current always flow between the two normal leads.
	In the full system, instead, due to the presence of the superconducting lead with a non-local coupling, the charge and the thermal current flow in different terminals enabling, thus,  nonlocal thermoelectricity.\cite{MazzaPRB2015a}
	
	In Fig.~\ref{fig.:linearRegime}, we compare the linear thermoelectric coefficient [panel (a)] and the thermal conductance [panel (b)] of this simplified model (dashed lines) with the results of the full calculation (solid lines) for different nonlocal couplings $\Gamma_S$ as a function of the 
	detuning. Here, we fix the free parameter $\alpha$ of the mapped quantum-dot model such that the linear current coefficient at $\Gamma_S=k_BT$ fits the one of the full model. The mapped quantum-dot model qualitatively captures the curve progression of the full computation.   When the nonlocal coupling is much larger than the temperature (blue lines) the behaviour exhibits two well separated resonances. At $\Gamma_S=4k_BT$ (green lines), the heat transport around  zero detuning is enhanced and this can be understood by the additive superposition of the contributions of both Lorentzians.  For lower values of the non-local coupling (red lines) the two resonances effectively merge and the behavior resembles that of a single resonance with a minimum in the thermal conductance at zero detuning. 
	The investigation (not shown) of the heat transport at the resonance $\Delta \epsilon= 0$  demonstrates that the maximum is obtained for $\Gamma_S=4 k_B T$, so this quantity can be used as an indirect way to measure the strength of the non-local coupling.   
	
	A few comments on the origin of the deviation between the full result (solid lines) and the simplified model (dashed lines) are in order. In the simplified model, the peaks around the resonances of the thermo-electrical coefficient, see Fig.~\ref{fig.:linearRegime}(a), are symmetric; this is not the case for the full results.
	The reason for the asymmetry is that the two resonances correspond to different Andreev levels which implies different energy-dependent weighting factors in front of the  Lorentzians.\cite{SplettstoesserPRB2007a} Similarly, for the thermal conductance, Fig.~\ref{fig.:linearRegime}(b), the simplified model underestimates the height of the  central peak. This can be again attributed to the fact that the central peak comes from the combined  action of the two Andreev resonances and not from independent resonances as naively  postulated in the simplified  model. The differences between the two models are a specific signature of the nature of the Andreev bound states in the DQD system with respect to standard QD resonances. 
	
	In order to better elecidate this peculiar signature of the proximity in our system, we need to go beyond the simple two-resonance model. In particular, we
	can derive the current and the heat current 
	for the case of symmetric detuning, $\epsilon_L=-\epsilon_R$, in the reduced Hilbert space that is constituted by the empty state, the singly occupied states, and the singlet state as discussed around Eq.~\eqref{eq:HeffReduced}.
	Under this assumptions, we find explicit expressions for the charge and the heat current, as reported in Appendix~\ref{app.:masterEq}. From those equations, we can easily compute the nonlocal linear response coefficients following the definitions given in Eq.~\eqref{linearcurrent} and Eq.~\eqref{linearheat}. One finds the linear response coefficients
	\begin{equation}
		\frac{\hbar L_{12}^{S}}{e_0 k_B} = -2 \frac{\Gamma_N}{k_B T} K(
		\Delta{\tilde\epsilon},
		\Delta{\tilde\epsilon},
		-\sqrt{2}{\tilde\Gamma_S}
		),
		\label{eq:LinCoeffsVSym_LS12}
	\end{equation}
	and
	\begin{equation}
		\frac{\hbar L_{22}^{R}}{k_B\Gamma_N} {=} K\Big(
		\Delta{\tilde\epsilon}^2 +\frac{5{\tilde\Gamma_S}^2}{2},
		\Delta{\tilde\epsilon}^2 +2{\tilde\Gamma_S}^2,
		-2\sqrt{2}{\tilde\Gamma_S}\Delta{\tilde\epsilon}
		\Big),
		\label{eq:LinCoeffsVSym_LR22}
	\end{equation}
	where ${\Delta\tilde\epsilon}=\Delta\epsilon/2k_BT$ and ${\tilde\Gamma_S}=\Gamma_S/2k_BT$ are dimensionless parameters.
	Both quantities can be written in terms of the same universal function
	\begin{equation*}
		K(x,y,z){=} \frac{
		x + y\cosh{\tilde\epsilon}\cosh\sqrt{2}{\tilde\Gamma_S}
		+ z\sinh{\tilde\epsilon}\sinh\sqrt{2}{\tilde\Gamma_S}
		}{
		3(
		\cosh{\tilde\epsilon}+\cosh\sqrt{2}{\tilde\Gamma_S}
		)(
		2\cosh{\tilde\epsilon}+\cosh\sqrt{2}{\tilde\Gamma_S}
		)
		}.
	\end{equation*}
	Furthermore, in the reduced model the non-local linear transport coefficient $L_{21}^{R}$ fulfills the standard Onsager relation $L_{21}^{R}= T L_{12}^S$. This indirectly supports that the linear regime is associated to a sort of nonlocal reversibility condition.
	In Fig~\ref{fig.:linearRegime}, we 
	show the transport coefficients of the reduced Hilbert space model with circles. In contrast to the mapped 
	single-quantum-dot model (dashed lines), they capture the asymmetry in the
	peak heights of the linear current [panel (a)], and explain better the enhanced 
	peaks of the linear heat current [panel (b)]. Moreover,
	the transport coefficients of the reduced model coincide with the ones of the full model (solid lines) in the case of a strong nonlocal 
	coupling $\Gamma_S\gg k_BT$ (blue case) without any free parameter.
	
	It is sometimes convenient to quantify the thermoelectricity in term of the Seebeck potential $\mu_S$. In Appendix A we compute the general formula for this quantity in Eq.~\eqref{eq.:SeebeckSymApprox} in the reduced Hilbert space model, for symmetric detuning $\epsilon_L=-\epsilon_R$ and small temperature gradient $\delta T$. Taking the limit of $\Gamma_S,\abs{\Delta\epsilon}\gg k_BT$, one can further simplify the expression for the Seebeck potential to
	\begin{equation}
		\label{eq.:Seebeck}
		\mu_{S\pm}\approx\big(\Delta\epsilon\mp\sqrt{2}\Gamma_S\big)\frac{\delta T}{4T},
	\end{equation}
	where the sign in $\mu_{S\pm}$ is simply determined by the sign of $\Delta\epsilon$. One can explain  the above result for the Seebeck potential again in terms of the mapped quantum-dot model where the thermoelectric effect is determined by two different resonances located at $\Delta E_\pm$. By definition, at the Seebeck potential, the thermocurrent generated by a temperature gradient vanishes. This is due to the fact that, at the addition 
	energies $\Delta E_\pm$, the electron distribution of the right normal lead 
	(having the tendency to push electrons into the superconductor) is identical to the hole 
	distribution on the left normal lead (having the tendency to pull electrons from the superconductor). Indeed the nonlocal thermocurrent is the resultant of this two competitive processes 
	governed by 
	the nonlocal Andreev bound state levels where the  quasi-particle state on one dot is coupled with quasi-hole state of the other, i.e. nonlocal particle-hole symmetry. 
	This implies that the condition to calculate the Seebeck 
	potential $\mu_S$, for a linear 
	temperature gradient $\delta T$, is given by the two equations
	$f_R(\Delta E_\pm)=1-f_L(-\Delta E_\mp)$, where the Fermi functions are computed at the equilibrium temperature $T=(T_R+T_L)$/2. The solution of these two conditions returns exactly the two results of Eq.~\eqref{eq.:Seebeck}, providing a  physical interpretation for the full formula given in Appendix A.

	\subsection{Nonlocal cooling}
	\begin{figure*}[t]
		\includegraphics{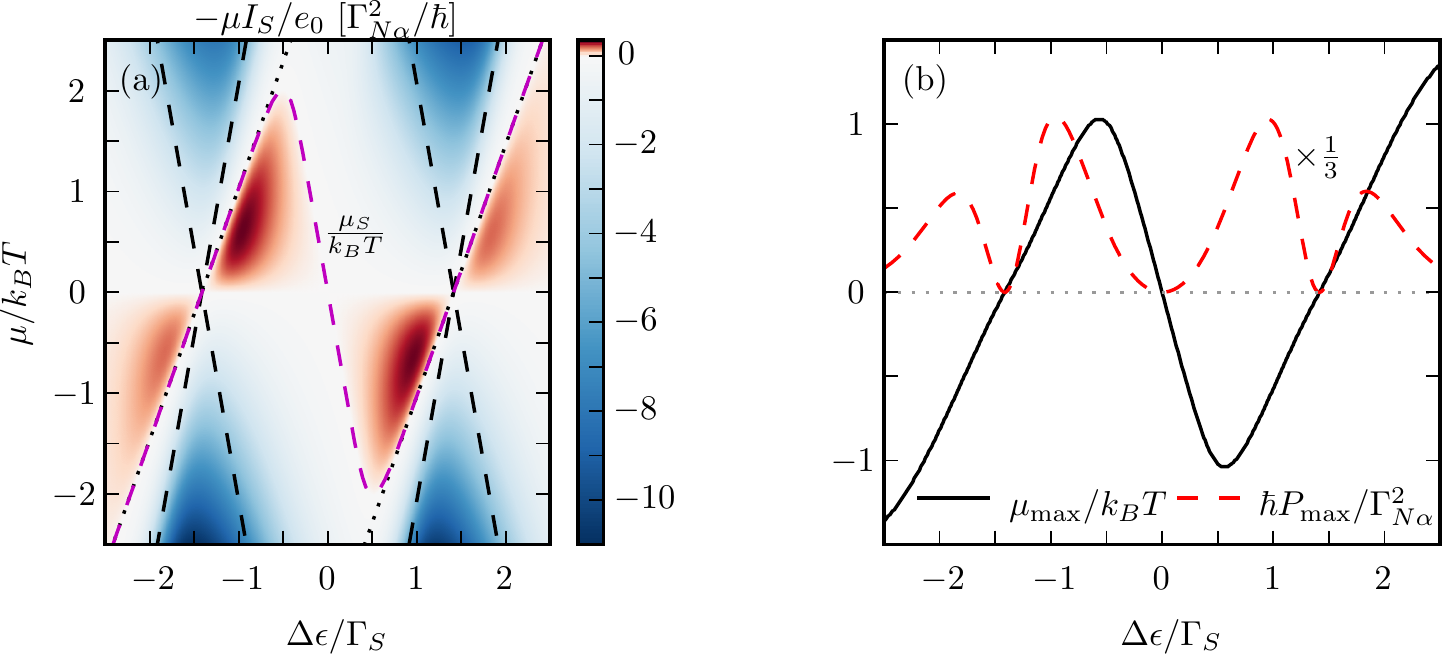}\caption{\label{fig.:maxPower}%
		(a) Thermopower $P=-\mu I_S/e_0$ %
		as a function of the detuning $\Delta\epsilon\equiv\epsilon_R-\epsilon_L$ and the chemical potential $\mu\equiv\mu_L=\mu_R$, 
		choosing the average level energy to be zero, i.e. $\epsilon_R+\epsilon_L=0$, and $k_B T_L= 3k_B T_R =15 \Gamma_{N\alpha}$. Other parameters are chosen as in Fig.~\ref{fig.:nonlocalCurrent}(a). The dashed black lines indicate where the addition energies $\Delta E_\pm$ are in resonance with 
		the chemical potential $\mu$, while the dotted lines correspond to the estimated nonlocal Seebeck potential
		$\mu_{S\pm}$ given in Eq.~\ref{eq.:Seebeck}. The dashed purple line corresponds to the condition \eqref{eq.:SeebeckSym} for the nonlocal Seebeck potential $\mu_S$.
		(b) Maximal power $P_{\textrm{max}} = \max_\mu P(\Delta\epsilon,\mu)$ (dashed line) and chemical potential $\mu_{\textrm{max}}$
		corresponding to the maximal power (solid line) for fixed detuning.
		}
	\end{figure*}
	
	We conclude this section with a final remark on the possibility to obtain a cooling cycle. 
	When a thermo-electrical device is operated near the reversibility condition, the thermo-electrical cycle can be inverted in order to get a cooling cycle.\cite{MazzaPRB2015a} Inspecting the level structure sketched in Fig.~\ref{fig.:sketchHeatingCooling}(b) one would expect nonlocal cooling at finite chemical potential.
	With cooling of a normal lead, we mean that electrons are either added 
	below its Fermi level or extracted above its Fermi level. 
	In particular, we expect that in the linear regime, by
	slightly moving the 
	chemical potential around the values $\mu_{S\pm}$, the nonlocal thermogenerator could turn into a nonlocal cooler and our thermoelectrical engine becomes a cooling device (Peltier cooling).
	
	In order to verify this mechanism, we firstly give a closer look to the heat current ${\dot Q}_R$ of Eq.~\eqref{eq.:currHeatSym},  linearizing the chemical potential $\mu\equiv\mu_S+\delta\mu_S$ around the nonlocal Seebeck potential $\mu_S$ of Eq.~\eqref{eq.:SeebeckSymApprox}. For simplicity, we consider equal temperatures, $T=T_L=T_R$, a condition for which nonlocal cooling is still possible. One finds that ${\dot Q}_R|_{\delta T=0}\approx L_{21}^{R} \delta\mu_S/e_0$, which shows clearly that the  heat flux changes its sign with the sign of $\delta\mu_S$. So, when the chemical potential crosses the nonlocal Seebeck potential, indeed, the system reverses the heat flux. In the presence of a small temperature gradient between the normal leads this heat-flux reversal corresponds to the conversion of a thermoelectrical generator into a Peltier cooler.
	
	This scenario is further supported by the behaviour of the 
	total electric power $P\equiv-\mu I_S/e_0$ generated around the nonlocal Seebeck potential $\mu_S$. Indeed, by linearizing in $\delta\mu_S$ the thermopower becomes $P\approx L_{12}^{S} \delta T\delta\mu_S/e_0$ where $\delta T$ is the temperature difference.  %
	Coherently, one sees  
	that the crossing of the nonlocal Seebeck potential, $\delta\mu_S\to-\delta\mu_S$, results in a sign change of the total power and, thus, entails a change between power generation to
	dissipation, as expected from consistency with general
	thermodynamical arguments for the scenario described so far.
	
	We will discuss the general behaviour of nonlocal thermoelectricity in more depth in Sec.~\ref{sec.:.powerCooling}, which treats the non-linear regime, and address, therein, quantitatively 
	the thermodynamical performances of the thermopower and the cooling. Furthermore, we will see that nonlocal cooling, indeed, sets in at $\mu\approx\mu_{S\pm}$ also well beyond the discussed linear regime.
	However before doing so, we discuss in the following section the effect of finite Coulomb interaction on the heat transport properties at the nonlocal resonance.

	\subsection{Effect of finite Coulomb interaction}
	
	Thus far, we have restricted our analysis to the case of infinite local Coulomb interaction in the QDs, so 
	that the  double occupation of the individual dots is forbidden.  Relaxing this condition
	and considering finite values for $U_\alpha$ opens up the possibility of a local exchange
	of Cooper pairs  between the superconductor and the
	system (both electrons in the Copper pair tunnel to/from the same dot).\footnote{In order to have a finite nonlocal coupling $\Gamma_S$ both the two local coupling $\Gamma_{S\alpha}$ with the superconductor need to be finite as $\Gamma_S\propto\sqrt{\Gamma_{SR}\Gamma_{SL}}$.}
	This includes the possibility to consider different virtual transitions involving a nonlocal
	resonance.  In this way, electrons can transfer energy between the normal
	leads via elastic cotunneling---in the subgap regime ($\abs{\Delta}\to\infty$) elastic cotunneling is not mediated by quasiparticles with energies larger then $\Delta$, but rather by  multiple 
	exchanges of local and nonlocal Cooper pairs triggered by $\Gamma_S$ and $\Gamma_{S\alpha}$. 
	Thus, the finite intradot Coulomb interaction can increase 
	the heat current $\dot Q_R$, while the current $I_S$ through the superconductor 
	remains unaffected. 
	In particular, an electron with spin $\sigma$ above the
	chemical potential of the left normal lead may tunnel with the rate
	$\Gamma_{NL}$ into the left dot and occupy the state $\ket{L\sigma}$. Then
	a Cooper-pair may split nonlocally with the coupling $\Gamma_S$ into the triply
	occupied state
	$\ket{tR\sigma}=d^\dag_{R\sigma}d^\dag_{L\uparrow}d^\dag_{L\downarrow}\ket{0}$
	followed by a local Cooper-pair recombination with the rate $\Gamma_{SL}$. Finally, the electron leaves the dot
	with the rate $\Gamma_{NR}$ via the right normal lead, heating up the right 
	lead. The process can also proceed differently with the local coupling operating before the nonlocal one.
	In this process the electron is effectively transferred to the state $\ket{R\sigma}$ of the right dot with no net current in the superconductor. This shows again that, due to nonlocality, the resonant behaviour of the heat does not necessarily affect the charge current. 
	The aforementioned mechanism can be  identified in the thermal transport at finite interaction such as in
	Fig.~\ref{fig.:finiteUEffect}(a), where the thermal transport coefficient 
	$L_{22}^R$ is shown as a function of the detuning $\Delta\epsilon$ and the average 
	temperature $T$. The thermal conductance $L_{22}^R$ describes how the heat current flows between the two normal terminals for $\mu=0$. 
	A remarkable feature is the narrow resonance at $\Delta\epsilon=0$.
	This resonance is absent for infinite local Coulomb interactions and its linewidth 
	increases with the scaling $\Gamma_{\alpha}/U_\alpha$, indicating its origin in the elastic cotunneling mechanism.
	
	In Fig.~\ref{fig.:finiteUEffect}(c) one can appreciate that for increasing
	intradot Coulomb interaction, the cotunneling peak becomes narrower, while its height remains unaffected. For this calculations the average quantum dot level has been chosen to be  
	$\epsilon\equiv(\epsilon_L+\epsilon_R)/2=\Gamma_S/3$ as since for $\epsilon=0$ the resonance is less pronounced [panel (b)].

	\section{\label{sec.:.powerCooling}Nonlocal thermoelectric power}
	
	So far, we have studied the transport coefficients for equal chemical potentials $\mu_R=\mu_L=\mu_{SC}=0$. In this case there is no power generation. For any circuital element, electrical work is performed if the charge carriers 
	gain potential energy by flowing against an increasing chemical potential.
	Therefore, while keeping the chemical potential of the superconductor at
	zero, for reference, we now consider the normal leads at non-zero values of $\mu$.  The
	corresponding generated work or thermopower $P\equiv-\mu I_S/e_0$ reflects
	the potential energy that an electron gains.\cite{ReyPRB2004a}  Upon increasing $\mu$ from a
	finite value, that still allows such counter-flow, to even larger ones, the flow of electrons will come at some specific value to a standstill.  Increasing $\mu$ further,
	will change the sign of the current.  Then, the thermoelectric element
	becomes dissipative and the electrons flow in the direction of the
	potential drop. In the inverted regime, a cooling effect can be also found before, at even higher voltages, where the fully dissipative regime dominates.
	
	Figure~\ref{fig.:maxPower}(a) shows the nonlocal thermopower as a
	function of the chemical potential and the detuning. As expected, the current becomes dissipative (negative power) for
	sufficiently large absolute values of $\mu$. 
	Nevertheless, there exist regions, namely the triangular
	ones in red, in which the Cooper pair splitter effectively generates
	positive thermopower and acts as a thermogenerator. 
	Furthermore, one can appreciate in this figure that the thermogenerator regime (red zones) is delimited by the nonlocal Seebeck formula Eq.~\eqref{eq.:SeebeckSym} (dashed purple line) derived in Appendix~\ref{app.:masterEq}, which can be approximated around the Andreev resonances by  Eq.~\eqref{eq.:Seebeck} (black dotted line). 
	
	In
	Fig.~\ref{fig.:maxPower}(b), we show the maximum generated power $P_{\rm
	max}$ (red dashed line) and the corresponding nonlocal Seebeck potential $\mu_{\rm
	max}$ (solid line) for which this maximum is obtained.
	The maximum generated power is relatively small, $P_{\rm max}\approx
	\Gamma_{N\alpha}^2/3\hbar$, and decaying for large detuning. 
	
	\begin{figure}[t]
		\includegraphics{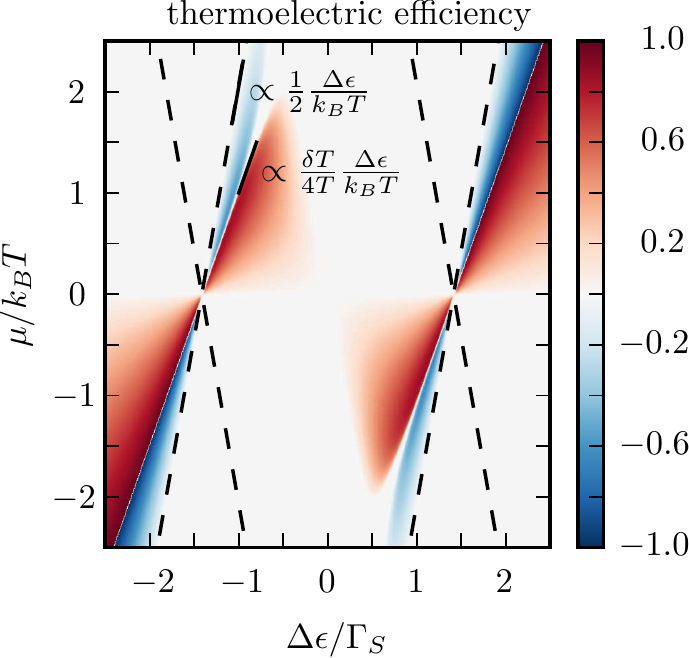}		\caption{\label{fig.:efficiency}%
		Power generation efficiency $P/(-{\dot Q}_L)$ (red) for $P>0$, and cooling efficiency ${\dot Q}_R/P$ (blue) for ${\dot Q}_R<0$ normalized by the 
		corresponding Carnot efficiencies, see Eq.~\eqref{eq.:rTherm}. In order to depict both efficiencies in the same plot, the latter is multiplied by a factor of $-1$.
		All parameters are as in Fig.~\ref{fig.:maxPower}.
		}
	\end{figure}
	
	Finally we discuss the thermoelectric efficiencies for power generation
	and cooling.
	For the nonlocal power generation, the efficiency $r_\mathrm{eng}=
	(P/\abs{{\dot Q}_L})/\eta_{C}$ is given by the power $P>0$
	generated in the system per extracted heat flow $-{\dot Q}_L>0$ from the
	warmer normal lead.\cite{BenentiPR2017a} It is normalized to the Carnot
	efficiency of a heat engine $\eta_{C}=1-T_R/T_L$, which is bounded
	between $0$ and $1$. Similarly,  the cooling power $r_{\rm fri}= \big({\dot
	Q}_R/P\big)/\eta_{\rm fri}$ is defined as the heat flow ${\dot Q}_R<0$
	extracted from the cold reservoir per absorbed power. We compare it with
	the ideal efficiency of a refrigerator $\eta_{\rm fri}=T_R/(T_L-T_R)$,
	which can be larger than one. 
	The combined efficiency 
	\begin{align}
			r_{\rm therm}=\begin{cases}
		r_{\rm eng}, & P>0\\
		-r_{\rm fri}, & {\dot Q}_R<0
			\end{cases}
	\label{eq.:rTherm}
	\end{align}
	is depicted in Fig.~\ref{fig.:efficiency} as a function of the detuning and the chemical potential,
	where  positive values (red) correspond to power generation and
	negative values (blue) to cooling of the right normal lead.  Close to the
	lines where the thermoelectric current vanishes, $\mu\propto(\delta T/4T)\Delta\epsilon$, see Eq.\eqref{eq.:Seebeck}, the system has a very high
	efficiency and represents an almost reversible thermoelectric generator (red
	shaded area). However, the power generated under this condition is
	rather small.  Therefore, as usual, there is a trade-off between high
	efficiency and high output.
	In particular at the nonlocal Seebeck potential, where the thermoelectric current 
	and the thermopower vanish, one generically expects that the thermoelectrical 
	machine becomes reversible and reaches the Carnot efficiency.
	
	The fact that the thermal machine operates nearly at the Carnot efficiency
	for some finite $\mu\neq0$, suggests that the system can become a cooling
	device. This happens,
	indeed, in the blue shaded region bounded by the nonlocal Andreev resonance
	at $\mu\propto\Delta\epsilon/2$, where the colder lead is further cooled
	due to nonlocal Cooper-pair tunneling.
	
	\section{\label{sec.:conclusions}Conclusions} 
	Nonlocal thermoelectric effects in a double-dot Cooper-pair splitter have been investigated. Thermoelectricity properties are determined by the non-local breaking of the particle-hole symmetry which is realized in the hybrid three-terminal structure in the presence of a temperature gradient. Intriguingly, we demonstrated that the superconductor can mediate coherent heat transfer between the normal leads.  The rich phenomenology can be easily interpreted in terms of a simple model consisting of two resonances located at the nonlocal Andreev bound state addition energies. However, this model has some limitations and the full model is required to get accurate results for the thermal transport. In particular, the Andreev nature of the resonances is reflected in a different energy dependence of those resonances. In comparison to the simple model, we predict an enhancement of the heat transferred between the normal lead at resonance for $\Gamma_S=4k_B T$. Finally, at the nonlocal resonance for finite Coulomb interaction an extra resonance is located in the heat transport as a consequence of virtual transitions to triple occupied states.  
	When applying a load between the normal leads and the superconducting one, the Cooper-pair splitter can perform work
	and convert heat current into electric current with nearly Carnot efficiency. The detuning can be used as control knob
	to turn the nonlocal power generator to a heat pump and cool the colder normal lead via nonlocal Cooper-pair tunneling.
	
	\begin{acknowledgments}
		When writing this paper, we became aware of a related work by Rafael S\'anchez et al., Ref.~\onlinecite{SanchezPRB2018a} and during the revision by Kirsanov et al. Ref.~\onlinecite{KirsanovArXiv2018a} and Pershoguba et al. Ref.~\onlinecite{PershogubArXiv2019a}. 
		
		R.H. acknowledges financial support from the Carl-Zeiss-Stiftung.
		This work was supported by the Spanish Ministry of Economy and
		Competitiveness via Grants MAT2017-86717-P and MAT2016-82015-REDT.
		W.B. acknowledges financial support from the DFG through SPP 2137 ``Spin Caloric Transport''.
		A.B. and F.G. acknowledge the European Research Council under the European Union's Seventh Framework Program (FP7/2007-2013)/ERC Grant agreement No.~615187-COMANCHE and the Tuscany Region under the FARFAS 2014 project SCIADRO for partial financial support. A.B. acknowledges the CNR-CONICET cooperation programme ``Energy conversion in quantum nanoscale hybrid devices'' and the Royal Society through the International Exchanges between the UK and Italy (grant IES R3 170054). M.G. acknowledges the hospitality of Scuola Normale Superiore, Pisa. 
	\end{acknowledgments}
	
	\appendix
	\section{\label{app.:masterEq}Current and heat current for strong nonlocal coupling} %
	In this section, we provide analytical expressions for the current, Eq.~\eqref{eq.:IAlpha}, and the 
	heat current, Eq.~\eqref{eq.:dotQ}, assuming symmetric detuning, $\epsilon_L=-\epsilon_R$, and equal 
	couplings to the normal leads, $\Gamma_N\equiv\Gamma_{N\alpha}$. As discussed in the main text, we 
	derive these quantities in the reduced Hilbert space spanned by the empty state, 
	the singly occupied states, and the singlet state which is a good approximation  at nonlocal resonance and for strong intradot Coulomb interactions $U_R,U_L\to\infty$. By using the general method presented in 
	Sec.~\ref{sec.:model}, 
	we find the expressions
	\begin{align}
		\begin{split}
	I_R
		&= \frac{e_0}{\hbar}\frac{N_I}{D}, \quad I_S = -2I_R, \\
	{\dot Q}_R
		&=-\frac{1}{\hbar}\frac{N_E}{D} +\frac{1}{e_0}\Big(\frac{\Delta\epsilon}{2} -\mu\Big) I_R,
		\end{split}
\label{eq.:currHeatSym}
	\end{align}
	with the numerator
	\begin{widetext}
		\begin{align}
	N_I ={ }&\Gamma_N\Big(\sum_{p,q}g_p^{p,q}-2\Big)\big[
		\big({\bar g}_+^{+,+} -g_-^{-,-}\big)\big({\bar g}_+^{+,-} -g_-^{-,+}\big) -9
	\big]
\intertext{of the current and the numerator of the energy current}
	N_E ={ }&\frac{9\Gamma_N\Gamma_S}{2\sqrt{2}}\sum_p p\big[
		g_-^{-,-p}g_+^{+,p} +\big(2+g_-^{-,p}g_+^{+,-p}\big)\big(
			g_-^{-,-p}+g_+^{+,p}
		\big)
	\big].
\intertext{The common denominator is given by}
	D ={ }&32+2\sum_{p,q} g_p^{p,q}\big[
		3+g_p^{p,q}\big(g_p^{p,-q}-{\bar g}_{-p}^{-p,q}\big)
	\big] \nonumber\\ 
	{ }&+\sum_p\big[ 
		10 g_p^{p,-}g_p^{p,+} +g_-^{-,p}\big(5g_+^{+,-p}-17g_+^{+,p}\big) 
		-5g_p^{p,p}g_p^{p,-p}\big(g_{-p}^{-p,-p}+g_{-p}^{-p,p}\big)
	\big].
		\end{align}
	\end{widetext}
	Here, $g_\alpha^{p,q} =\big\{ 
	1+\exp\big[(p\frac{\Delta\epsilon}{2} +q\frac{\Gamma_S}{\sqrt{2}}-\mu)/k_BT_\alpha\big] 
	\big\}^{-1}$
	compactly denotes the Fermi function evaluated at the Andreev bound state addition energies with $p,q\in\{-1,1\}$, 
	and ${\bar g}_\alpha^{p,q}\equiv1-g_\alpha^{p,q}$. For the sake of a compact notation, 
	we also identified the terminals $\alpha=L,R$ with the values $\alpha=\mp$. The corresponding 
	current and heat current on the left normal lead follow from the mutual
	replacement $\{\epsilon_L\leftrightarrow\epsilon_R, T_L\leftrightarrow T_R\}$. 
	We note that $N_I$ and $D$ are unaffected under this transformation
	since the currents through the normal leads are identical, 
	$I_L=I_R$. %
	On the contrary, $N_E$ experiences a change in sign leading to the energy conservation 
	${\dot Q}_L+{\dot Q}_R=\mu I_S$.
	
	A closer inspection of the numerator $N_I$ reveals, that the current through the superconductor 
	only vanishes if the term in its first parenthesis nullifies, since the denominator is always finite.
	This gives us the condition
	\begin{align}
	2 &=\sum_{p,q}g_p^{p,q}\big|_{\mu=\mu_S} \label{eq.:SeebeckSym}
	\end{align}
	which defines implicitly the nonlocal Seebeck potential $\mu_S$. 
	Figure~\ref{fig.:maxPower} of the main text visualizes that this expression $\mu_S$ (dashed purple line)
	asymptotically approaches the estimates $\mu_{S\pm}$ (dotted black lines), Eq.~\eqref{eq.:Seebeck}, of the mapped single-quantum-dot model. This asymptotically behaviour, approximated as
	\begin{equation}
		\mu_S\approx \Big[
		\Delta\epsilon -\frac{\sqrt{2}\Gamma_S
		\sinh\big(\frac{\Delta\epsilon}{2 k_BT}\big)\sinh\big(\frac{\Gamma_S}{\sqrt{2}k_BT}\big)
		}{
		1+\cosh\big(\frac{\Delta\epsilon}{2 k_BT}\big)\cosh\big(\frac{\Gamma_S}{\sqrt{2}k_BT}\big)
		}
		\Big] \frac{\delta T}{4T},
		\label{eq.:SeebeckSymApprox}
	\end{equation}
	follows from the linearisation of Eq.~\eqref{eq.:SeebeckSym} in $\mu_S$ and the temperature 
	difference $\delta T$. In the limit $\Gamma_S,\abs{\Delta\epsilon}\gg k_BT$, this reduces 
	for $\sgn(\Delta\epsilon)=\pm1$ to the branches $\mu_{S\pm}$, respectively.


\begin{thebibliography}{97}%
	\makeatletter
	\providecommand \@ifxundefined [1]{%
	 \@ifx{#1\undefined}
	}%
	\providecommand \@ifnum [1]{%
	 \ifnum #1\expandafter \@firstoftwo
	 \else \expandafter \@secondoftwo
	 \fi
	}%
	\providecommand \@ifx [1]{%
	 \ifx #1\expandafter \@firstoftwo
	 \else \expandafter \@secondoftwo
	 \fi
	}%
	\providecommand \natexlab [1]{#1}%
	\providecommand \enquote  [1]{``#1''}%
	\providecommand \bibnamefont  [1]{#1}%
	\providecommand \bibfnamefont [1]{#1}%
	\providecommand \citenamefont [1]{#1}%
	\providecommand \href@noop [0]{\@secondoftwo}%
	\providecommand \href [0]{\begingroup \@sanitize@url \@href}%
	\providecommand \@href[1]{\@@startlink{#1}\@@href}%
	\providecommand \@@href[1]{\endgroup#1\@@endlink}%
	\providecommand \@sanitize@url [0]{\catcode `\\12\catcode `\$12\catcode
	  `\&12\catcode `\#12\catcode `\^12\catcode `\_12\catcode `\%12\relax}%
	\providecommand \@@startlink[1]{}%
	\providecommand \@@endlink[0]{}%
	\providecommand \url  [0]{\begingroup\@sanitize@url \@url }%
	\providecommand \@url [1]{\endgroup\@href {#1}{\urlprefix }}%
	\providecommand \urlprefix  [0]{URL }%
	\providecommand \Eprint [0]{\href }%
	\providecommand \doibase [0]{http://dx.doi.org/}%
	\providecommand \selectlanguage [0]{\@gobble}%
	\providecommand \bibinfo  [0]{\@secondoftwo}%
	\providecommand \bibfield  [0]{\@secondoftwo}%
	\providecommand \translation [1]{[#1]}%
	\providecommand \BibitemOpen [0]{}%
	\providecommand \bibitemStop [0]{}%
	\providecommand \bibitemNoStop [0]{.\EOS\space}%
	\providecommand \EOS [0]{\spacefactor3000\relax}%
	\providecommand \BibitemShut  [1]{\csname bibitem#1\endcsname}%
	\let\auto@bib@innerbib\@empty
	\bibitem [{\citenamefont {Lesovik}\ \emph {et~al.}(2001)\citenamefont
	  {Lesovik}, \citenamefont {Martin},\ and\ \citenamefont
	  {Blatter}}]{LesovikEPJB2001a}%
	  \BibitemOpen
	  \bibfield  {author} {\bibinfo {author} {\bibfnamefont {G.~B.}\ \bibnamefont
	  {Lesovik}}, \bibinfo {author} {\bibfnamefont {T.}~\bibnamefont {Martin}}, \
	  and\ \bibinfo {author} {\bibfnamefont {G.}~\bibnamefont {Blatter}},\
	  }\href@noop {} {\bibfield  {journal} {\bibinfo  {journal} {Eur. Phys. J. B}\
	  }\textbf {\bibinfo {volume} {24}},\ \bibinfo {pages} {287} (\bibinfo {year}
	  {2001})}\BibitemShut {NoStop}%
	\bibitem [{\citenamefont {Russo}\ \emph {et~al.}(2005)\citenamefont {Russo},
	  \citenamefont {Kroug}, \citenamefont {Klapwijk},\ and\ \citenamefont
	  {Morpurgo}}]{RussoPRL2005a}%
	  \BibitemOpen
	  \bibfield  {author} {\bibinfo {author} {\bibfnamefont {S.}~\bibnamefont
	  {Russo}}, \bibinfo {author} {\bibfnamefont {M.}~\bibnamefont {Kroug}},
	  \bibinfo {author} {\bibfnamefont {T.~M.}\ \bibnamefont {Klapwijk}}, \ and\
	  \bibinfo {author} {\bibfnamefont {A.~F.}\ \bibnamefont {Morpurgo}},\
	  }\href@noop {} {\bibfield  {journal} {\bibinfo  {journal} {Phys. Rev. Lett.}\
	  }\textbf {\bibinfo {volume} {95}},\ \bibinfo {pages} {027002} (\bibinfo
	  {year} {2005})}\BibitemShut {NoStop}%
	\bibitem [{\citenamefont {Mart\'in-Rodero}\ and\ \citenamefont
	  {Yeyati}(2011)}]{Martin-RoderoAP2011a}%
	  \BibitemOpen
	  \bibfield  {author} {\bibinfo {author} {\bibfnamefont {A.}~\bibnamefont
	  {Mart\'in-Rodero}}\ and\ \bibinfo {author} {\bibfnamefont {A.~L.}\
	  \bibnamefont {Yeyati}},\ }\href@noop {} {\bibfield  {journal} {\bibinfo
	  {journal} {Adv. Phys.}\ }\textbf {\bibinfo {volume} {60}},\ \bibinfo {pages}
	  {899} (\bibinfo {year} {2011})}\BibitemShut {NoStop}%
	\bibitem [{\citenamefont {Roddaro}\ \emph {et~al.}(2011)\citenamefont
	  {Roddaro}, \citenamefont {Pescaglini}, \citenamefont {Ercolani},
	  \citenamefont {Sorba}, \citenamefont {Giazotto},\ and\ \citenamefont
	  {Beltram}}]{RoddaroNanoR2011a}%
	  \BibitemOpen
	  \bibfield  {author} {\bibinfo {author} {\bibfnamefont {S.}~\bibnamefont
	  {Roddaro}}, \bibinfo {author} {\bibfnamefont {A.}~\bibnamefont {Pescaglini}},
	  \bibinfo {author} {\bibfnamefont {D.}~\bibnamefont {Ercolani}}, \bibinfo
	  {author} {\bibfnamefont {L.}~\bibnamefont {Sorba}}, \bibinfo {author}
	  {\bibfnamefont {F.}~\bibnamefont {Giazotto}}, \ and\ \bibinfo {author}
	  {\bibfnamefont {F.}~\bibnamefont {Beltram}},\ }\href@noop {} {\bibfield
	  {journal} {\bibinfo  {journal} {Nano Res.}\ }\textbf {\bibinfo {volume}
	  {4}},\ \bibinfo {pages} {259} (\bibinfo {year} {2011})}\BibitemShut {NoStop}%
	\bibitem [{\citenamefont {Giazotto}\ \emph {et~al.}(2011)\citenamefont
	  {Giazotto}, \citenamefont {Spathis}, \citenamefont {Roddaro}, \citenamefont
	  {Biswas}, \citenamefont {Taddei}, \citenamefont {Governale},\ and\
	  \citenamefont {Sorba}}]{GiazottoNatPhys2011a}%
	  \BibitemOpen
	  \bibfield  {author} {\bibinfo {author} {\bibfnamefont {F.}~\bibnamefont
	  {Giazotto}}, \bibinfo {author} {\bibfnamefont {P.}~\bibnamefont {Spathis}},
	  \bibinfo {author} {\bibfnamefont {S.}~\bibnamefont {Roddaro}}, \bibinfo
	  {author} {\bibfnamefont {S.}~\bibnamefont {Biswas}}, \bibinfo {author}
	  {\bibfnamefont {F.}~\bibnamefont {Taddei}}, \bibinfo {author} {\bibfnamefont
	  {M.}~\bibnamefont {Governale}}, \ and\ \bibinfo {author} {\bibfnamefont
	  {L.}~\bibnamefont {Sorba}},\ }\href@noop {} {\bibfield  {journal} {\bibinfo
	  {journal} {Nat. Phys.}\ }\textbf {\bibinfo {volume} {7}},\ \bibinfo {pages}
	  {857} (\bibinfo {year} {2011})}\BibitemShut {NoStop}%
	\bibitem [{\citenamefont {Schindele}\ \emph {et~al.}(2012)\citenamefont
	  {Schindele}, \citenamefont {Baumgartner},\ and\ \citenamefont
	  {Sch\"onenberger}}]{SchindelePRL2012a}%
	  \BibitemOpen
	  \bibfield  {author} {\bibinfo {author} {\bibfnamefont {J.}~\bibnamefont
	  {Schindele}}, \bibinfo {author} {\bibfnamefont {A.}~\bibnamefont
	  {Baumgartner}}, \ and\ \bibinfo {author} {\bibfnamefont {C.}~\bibnamefont
	  {Sch\"onenberger}},\ }\href@noop {} {\bibfield  {journal} {\bibinfo
	  {journal} {Phys. Rev. Lett.}\ }\textbf {\bibinfo {volume} {109}},\ \bibinfo
	  {pages} {157002} (\bibinfo {year} {2012})}\BibitemShut {NoStop}%
	\bibitem [{\citenamefont {Romeo}\ \emph {et~al.}(2012)\citenamefont {Romeo},
	  \citenamefont {Roddaro}, \citenamefont {Pitanti}, \citenamefont {Ercolani},
	  \citenamefont {Sorba},\ and\ \citenamefont {Beltram}}]{RomeoNanoL2012a}%
	  \BibitemOpen
	  \bibfield  {author} {\bibinfo {author} {\bibfnamefont {L.}~\bibnamefont
	  {Romeo}}, \bibinfo {author} {\bibfnamefont {S.}~\bibnamefont {Roddaro}},
	  \bibinfo {author} {\bibfnamefont {A.}~\bibnamefont {Pitanti}}, \bibinfo
	  {author} {\bibfnamefont {D.}~\bibnamefont {Ercolani}}, \bibinfo {author}
	  {\bibfnamefont {L.}~\bibnamefont {Sorba}}, \ and\ \bibinfo {author}
	  {\bibfnamefont {F.}~\bibnamefont {Beltram}},\ }\href@noop {} {\bibfield
	  {journal} {\bibinfo  {journal} {Nano Lett.}\ }\textbf {\bibinfo {volume}
	  {12}},\ \bibinfo {pages} {4490} (\bibinfo {year} {2012})}\BibitemShut
	  {NoStop}%
	\bibitem [{\citenamefont {Das}\ \emph {et~al.}(2012)\citenamefont {Das},
	  \citenamefont {Ronen}, \citenamefont {Heiblum}, \citenamefont {Mahalu},
	  \citenamefont {Kretinin},\ and\ \citenamefont {Shtrikman}}]{DasNatureC2012a}%
	  \BibitemOpen
	  \bibfield  {author} {\bibinfo {author} {\bibfnamefont {A.}~\bibnamefont
	  {Das}}, \bibinfo {author} {\bibfnamefont {Y.}~\bibnamefont {Ronen}}, \bibinfo
	  {author} {\bibfnamefont {M.}~\bibnamefont {Heiblum}}, \bibinfo {author}
	  {\bibfnamefont {D.}~\bibnamefont {Mahalu}}, \bibinfo {author} {\bibfnamefont
	  {A.~V.}\ \bibnamefont {Kretinin}}, \ and\ \bibinfo {author} {\bibfnamefont
	  {H.}~\bibnamefont {Shtrikman}},\ }\href@noop {} {\bibfield  {journal}
	  {\bibinfo  {journal} {Nat. Commun.}\ }\textbf {\bibinfo {volume} {3}},\
	  \bibinfo {pages} {1165} (\bibinfo {year} {2012})}\BibitemShut {NoStop}%
	\bibitem [{\citenamefont {Braunecker}\ \emph {et~al.}(2013)\citenamefont
	  {Braunecker}, \citenamefont {Burset},\ and\ \citenamefont
	  {Levy~Yeyati}}]{BrauneckerPRL2013a}%
	  \BibitemOpen
	  \bibfield  {author} {\bibinfo {author} {\bibfnamefont {B.}~\bibnamefont
	  {Braunecker}}, \bibinfo {author} {\bibfnamefont {P.}~\bibnamefont {Burset}},
	  \ and\ \bibinfo {author} {\bibfnamefont {A.}~\bibnamefont {Levy~Yeyati}},\
	  }\href@noop {} {\bibfield  {journal} {\bibinfo  {journal} {Phys. Rev. Lett.}\
	  }\textbf {\bibinfo {volume} {111}},\ \bibinfo {pages} {136806} (\bibinfo
	  {year} {2013})}\BibitemShut {NoStop}%
	\bibitem [{\citenamefont {Rossella}\ \emph {et~al.}(2014)\citenamefont
	  {Rossella}, \citenamefont {Bertoni}, \citenamefont {Ercolani}, \citenamefont
	  {Rontani}, \citenamefont {Sorba}, \citenamefont {Beltram},\ and\
	  \citenamefont {Roddaro}}]{RossellaNatNano2014a}%
	  \BibitemOpen
	  \bibfield  {author} {\bibinfo {author} {\bibfnamefont {F.}~\bibnamefont
	  {Rossella}}, \bibinfo {author} {\bibfnamefont {A.}~\bibnamefont {Bertoni}},
	  \bibinfo {author} {\bibfnamefont {D.}~\bibnamefont {Ercolani}}, \bibinfo
	  {author} {\bibfnamefont {M.}~\bibnamefont {Rontani}}, \bibinfo {author}
	  {\bibfnamefont {L.}~\bibnamefont {Sorba}}, \bibinfo {author} {\bibfnamefont
	  {F.}~\bibnamefont {Beltram}}, \ and\ \bibinfo {author} {\bibfnamefont
	  {S.}~\bibnamefont {Roddaro}},\ }\href@noop {} {\bibfield  {journal} {\bibinfo
	   {journal} {Nat. Nanotechnol.}\ }\textbf {\bibinfo {volume} {9}},\ \bibinfo
	  {pages} {997} (\bibinfo {year} {2014})}\BibitemShut {NoStop}%
	\bibitem [{\citenamefont {Choi}(2014)}]{ChoiPRB2014a}%
	  \BibitemOpen
	  \bibfield  {author} {\bibinfo {author} {\bibfnamefont {M.-S.}\ \bibnamefont
	  {Choi}},\ }\href@noop {} {\bibfield  {journal} {\bibinfo  {journal} {Phys.
	  Rev. B}\ }\textbf {\bibinfo {volume} {89}},\ \bibinfo {pages} {045137}
	  (\bibinfo {year} {2014})}\BibitemShut {NoStop}%
	\bibitem [{\citenamefont {Sato}\ and\ \citenamefont
	  {Tserkovnyak}(2014)}]{SatoPRB2014a}%
	  \BibitemOpen
	  \bibfield  {author} {\bibinfo {author} {\bibfnamefont {K.}~\bibnamefont
	  {Sato}}\ and\ \bibinfo {author} {\bibfnamefont {Y.}~\bibnamefont
	  {Tserkovnyak}},\ }\href@noop {} {\bibfield  {journal} {\bibinfo  {journal}
	  {Phys. Rev. B}\ }\textbf {\bibinfo {volume} {90}},\ \bibinfo {pages} {045419}
	  (\bibinfo {year} {2014})}\BibitemShut {NoStop}%
	\bibitem [{\citenamefont {Deacon}\ \emph {et~al.}(2015)\citenamefont {Deacon},
	  \citenamefont {Oiwa}, \citenamefont {Sailer}, \citenamefont {Baba},
	  \citenamefont {Kanai}, \citenamefont {Shibata}, \citenamefont {Hirakawa},\
	  and\ \citenamefont {Tarucha}}]{DeaconNatureC2015a}%
	  \BibitemOpen
	  \bibfield  {author} {\bibinfo {author} {\bibfnamefont {R.~S.}\ \bibnamefont
	  {Deacon}}, \bibinfo {author} {\bibfnamefont {A.}~\bibnamefont {Oiwa}},
	  \bibinfo {author} {\bibfnamefont {J.}~\bibnamefont {Sailer}}, \bibinfo
	  {author} {\bibfnamefont {S.}~\bibnamefont {Baba}}, \bibinfo {author}
	  {\bibfnamefont {Y.}~\bibnamefont {Kanai}}, \bibinfo {author} {\bibfnamefont
	  {K.}~\bibnamefont {Shibata}}, \bibinfo {author} {\bibfnamefont
	  {K.}~\bibnamefont {Hirakawa}}, \ and\ \bibinfo {author} {\bibfnamefont
	  {S.}~\bibnamefont {Tarucha}},\ }\href@noop {} {\bibfield  {journal} {\bibinfo
	   {journal} {Nat. Commun.}\ }\textbf {\bibinfo {volume} {6}},\ \bibinfo
	  {pages} {7446} (\bibinfo {year} {2015})}\BibitemShut {NoStop}%
	\bibitem [{\citenamefont {Marra}\ \emph {et~al.}(2016)\citenamefont {Marra},
	  \citenamefont {Citro},\ and\ \citenamefont {Braggio}}]{MarraPRB2016a}%
	  \BibitemOpen
	  \bibfield  {author} {\bibinfo {author} {\bibfnamefont {P.}~\bibnamefont
	  {Marra}}, \bibinfo {author} {\bibfnamefont {R.}~\bibnamefont {Citro}}, \ and\
	  \bibinfo {author} {\bibfnamefont {A.}~\bibnamefont {Braggio}},\ }\href@noop
	  {} {\bibfield  {journal} {\bibinfo  {journal} {Phys. Rev. B}\ }\textbf
	  {\bibinfo {volume} {93}},\ \bibinfo {pages} {220507} (\bibinfo {year}
	  {2016})}\BibitemShut {NoStop}%
	\bibitem [{\citenamefont {Tiira}\ \emph {et~al.}(2017)\citenamefont {Tiira},
	  \citenamefont {Strambini}, \citenamefont {Amado}, \citenamefont {Roddaro},
	  \citenamefont {San-Jose}, \citenamefont {Aguado}, \citenamefont {Bergeret},
	  \citenamefont {Ercolani}, \citenamefont {Sorba},\ and\ \citenamefont
	  {Giazotto}}]{TiiraNatureC2017a}%
	  \BibitemOpen
	  \bibfield  {author} {\bibinfo {author} {\bibfnamefont {J.}~\bibnamefont
	  {Tiira}}, \bibinfo {author} {\bibfnamefont {E.}~\bibnamefont {Strambini}},
	  \bibinfo {author} {\bibfnamefont {M.}~\bibnamefont {Amado}}, \bibinfo
	  {author} {\bibfnamefont {S.}~\bibnamefont {Roddaro}}, \bibinfo {author}
	  {\bibfnamefont {P.}~\bibnamefont {San-Jose}}, \bibinfo {author}
	  {\bibfnamefont {R.}~\bibnamefont {Aguado}}, \bibinfo {author} {\bibfnamefont
	  {F.~S.}\ \bibnamefont {Bergeret}}, \bibinfo {author} {\bibfnamefont
	  {D.}~\bibnamefont {Ercolani}}, \bibinfo {author} {\bibfnamefont
	  {L.}~\bibnamefont {Sorba}}, \ and\ \bibinfo {author} {\bibfnamefont
	  {F.}~\bibnamefont {Giazotto}},\ }\href@noop {} {\bibfield  {journal}
	  {\bibinfo  {journal} {Nat. Commun.}\ }\textbf {\bibinfo {volume} {8}},\
	  \bibinfo {pages} {14984} (\bibinfo {year} {2017})}\BibitemShut {NoStop}%
	\bibitem [{\citenamefont {Blasi}\ \emph {et~al.}(2018)\citenamefont {Blasi},
	  \citenamefont {Taddei}, \citenamefont {Giovannetti},\ and\ \citenamefont
	  {Braggio}}]{BlasiArXiv2018a}%
	  \BibitemOpen
	  \bibfield  {author} {\bibinfo {author} {\bibfnamefont {G.}~\bibnamefont
	  {Blasi}}, \bibinfo {author} {\bibfnamefont {F.}~\bibnamefont {Taddei}},
	  \bibinfo {author} {\bibfnamefont {V.}~\bibnamefont {Giovannetti}}, \ and\
	  \bibinfo {author} {\bibfnamefont {A.}~\bibnamefont {Braggio}},\ }\href@noop
	  {} {\bibfield  {journal} {\bibinfo  {journal} {arXiv:1808.09709}\ } (\bibinfo
	  {year} {2018})}\BibitemShut {NoStop}%
	\bibitem [{\citenamefont {Linder}\ and\ \citenamefont
	  {Robinson}(2015)}]{LinderNatPhys2015a}%
	  \BibitemOpen
	  \bibfield  {author} {\bibinfo {author} {\bibfnamefont {J.}~\bibnamefont
	  {Linder}}\ and\ \bibinfo {author} {\bibfnamefont {J.~W.~A.}\ \bibnamefont
	  {Robinson}},\ }\href@noop {} {\bibfield  {journal} {\bibinfo  {journal} {Nat.
	  Phys.}\ }\textbf {\bibinfo {volume} {11}},\ \bibinfo {pages} {307} (\bibinfo
	  {year} {2015})}\BibitemShut {NoStop}%
	\bibitem [{\citenamefont {Monroe}(2002)}]{MonroeNature2002a}%
	  \BibitemOpen
	  \bibfield  {author} {\bibinfo {author} {\bibfnamefont {C.}~\bibnamefont
	  {Monroe}},\ }\href@noop {} {\bibfield  {journal} {\bibinfo  {journal}
	  {Nature}\ }\textbf {\bibinfo {volume} {416}},\ \bibinfo {pages} {238}
	  (\bibinfo {year} {2002})}\BibitemShut {NoStop}%
	\bibitem [{\citenamefont {Ladd}\ \emph {et~al.}(2010)\citenamefont {Ladd},
	  \citenamefont {Jelezko}, \citenamefont {Laflamme}, \citenamefont {Nakamura},
	  \citenamefont {Monroe},\ and\ \citenamefont {O'Brien}}]{LaddNature2010a}%
	  \BibitemOpen
	  \bibfield  {author} {\bibinfo {author} {\bibfnamefont {T.~D.}\ \bibnamefont
	  {Ladd}}, \bibinfo {author} {\bibfnamefont {F.}~\bibnamefont {Jelezko}},
	  \bibinfo {author} {\bibfnamefont {R.}~\bibnamefont {Laflamme}}, \bibinfo
	  {author} {\bibfnamefont {Y.}~\bibnamefont {Nakamura}}, \bibinfo {author}
	  {\bibfnamefont {C.}~\bibnamefont {Monroe}}, \ and\ \bibinfo {author}
	  {\bibfnamefont {J.~L.}\ \bibnamefont {O'Brien}},\ }\href@noop {} {\bibfield
	  {journal} {\bibinfo  {journal} {Nature}\ }\textbf {\bibinfo {volume} {464}},\
	  \bibinfo {pages} {45} (\bibinfo {year} {2010})}\BibitemShut {NoStop}%
	\bibitem [{\citenamefont {Beckmann}\ \emph {et~al.}(2004)\citenamefont
	  {Beckmann}, \citenamefont {Weber},\ and\ \citenamefont
	  {v.~L\"ohneysen}}]{BeckmannPRL2004a}%
	  \BibitemOpen
	  \bibfield  {author} {\bibinfo {author} {\bibfnamefont {D.}~\bibnamefont
	  {Beckmann}}, \bibinfo {author} {\bibfnamefont {H.~B.}\ \bibnamefont {Weber}},
	  \ and\ \bibinfo {author} {\bibfnamefont {H.}~\bibnamefont {v.~L\"ohneysen}},\
	  }\href@noop {} {\bibfield  {journal} {\bibinfo  {journal} {Phys. Rev. Lett.}\
	  }\textbf {\bibinfo {volume} {93}},\ \bibinfo {pages} {197003} (\bibinfo
	  {year} {2004})}\BibitemShut {NoStop}%
	\bibitem [{\citenamefont {Hofstetter}\ \emph {et~al.}(2010)\citenamefont
	  {Hofstetter}, \citenamefont {Geresdi}, \citenamefont {Aagesen}, \citenamefont
	  {Nyg\aa{}rd}, \citenamefont {Sch\"onenberger},\ and\ \citenamefont
	  {Csonka}}]{HofstetterPRL2010a}%
	  \BibitemOpen
	  \bibfield  {author} {\bibinfo {author} {\bibfnamefont {L.}~\bibnamefont
	  {Hofstetter}}, \bibinfo {author} {\bibfnamefont {A.}~\bibnamefont {Geresdi}},
	  \bibinfo {author} {\bibfnamefont {M.}~\bibnamefont {Aagesen}}, \bibinfo
	  {author} {\bibfnamefont {J.}~\bibnamefont {Nyg\aa{}rd}}, \bibinfo {author}
	  {\bibfnamefont {C.}~\bibnamefont {Sch\"onenberger}}, \ and\ \bibinfo {author}
	  {\bibfnamefont {S.}~\bibnamefont {Csonka}},\ }\href@noop {} {\bibfield
	  {journal} {\bibinfo  {journal} {Phys. Rev. Lett.}\ }\textbf {\bibinfo
	  {volume} {104}},\ \bibinfo {pages} {246804} (\bibinfo {year}
	  {2010})}\BibitemShut {NoStop}%
	\bibitem [{\citenamefont {Trocha}\ and\ \citenamefont
	  {Weymann}(2015)}]{TrochaPRB2015a}%
	  \BibitemOpen
	  \bibfield  {author} {\bibinfo {author} {\bibfnamefont {P.}~\bibnamefont
	  {Trocha}}\ and\ \bibinfo {author} {\bibfnamefont {I.}~\bibnamefont
	  {Weymann}},\ }\href@noop {} {\bibfield  {journal} {\bibinfo  {journal} {Phys.
	  Rev. B}\ }\textbf {\bibinfo {volume} {91}},\ \bibinfo {pages} {235424}
	  (\bibinfo {year} {2015})}\BibitemShut {NoStop}%
	\bibitem [{\citenamefont {Wrze\'sniewski}\ \emph {et~al.}(2017)\citenamefont
	  {Wrze\'sniewski}, \citenamefont {Trocha},\ and\ \citenamefont
	  {Weymann}}]{WrzesniewskiJP2017a}%
	  \BibitemOpen
	  \bibfield  {author} {\bibinfo {author} {\bibfnamefont {K.}~\bibnamefont
	  {Wrze\'sniewski}}, \bibinfo {author} {\bibfnamefont {P.}~\bibnamefont
	  {Trocha}}, \ and\ \bibinfo {author} {\bibfnamefont {I.}~\bibnamefont
	  {Weymann}},\ }\href@noop {} {\bibfield  {journal} {\bibinfo  {journal} {J.
	  Phys.: Condens. Matter}\ }\textbf {\bibinfo {volume} {29}},\ \bibinfo {pages}
	  {195302} (\bibinfo {year} {2017})}\BibitemShut {NoStop}%
	\bibitem [{\citenamefont {Bocian}\ \emph {et~al.}(2018)\citenamefont {Bocian},
	  \citenamefont {Rudzi\ifmmode~\acute{n}\else \'{n}\fi{}ski},\ and\
	  \citenamefont {Weymann}}]{BocianPRB2018a}%
	  \BibitemOpen
	  \bibfield  {author} {\bibinfo {author} {\bibfnamefont {K.}~\bibnamefont
	  {Bocian}}, \bibinfo {author} {\bibfnamefont {W.}~\bibnamefont
	  {Rudzi\ifmmode~\acute{n}\else \'{n}\fi{}ski}}, \ and\ \bibinfo {author}
	  {\bibfnamefont {I.}~\bibnamefont {Weymann}},\ }\href@noop {} {\bibfield
	  {journal} {\bibinfo  {journal} {Phys. Rev. B}\ }\textbf {\bibinfo {volume}
	  {97}},\ \bibinfo {pages} {195441} (\bibinfo {year} {2018})}\BibitemShut
	  {NoStop}%
	\bibitem [{\citenamefont {Choi}\ \emph {et~al.}(2000)\citenamefont {Choi},
	  \citenamefont {Bruder},\ and\ \citenamefont {Loss}}]{ChoiPRB2000a}%
	  \BibitemOpen
	  \bibfield  {author} {\bibinfo {author} {\bibfnamefont {M.-S.}\ \bibnamefont
	  {Choi}}, \bibinfo {author} {\bibfnamefont {C.}~\bibnamefont {Bruder}}, \ and\
	  \bibinfo {author} {\bibfnamefont {D.}~\bibnamefont {Loss}},\ }\href@noop {}
	  {\bibfield  {journal} {\bibinfo  {journal} {Phys. Rev. B}\ }\textbf {\bibinfo
	  {volume} {62}},\ \bibinfo {pages} {13569} (\bibinfo {year}
	  {2000})}\BibitemShut {NoStop}%
	\bibitem [{\citenamefont {Recher}\ \emph {et~al.}(2001)\citenamefont {Recher},
	  \citenamefont {Sukhorukov},\ and\ \citenamefont {Loss}}]{RecherPRB2001a}%
	  \BibitemOpen
	  \bibfield  {author} {\bibinfo {author} {\bibfnamefont {P.}~\bibnamefont
	  {Recher}}, \bibinfo {author} {\bibfnamefont {E.~V.}\ \bibnamefont
	  {Sukhorukov}}, \ and\ \bibinfo {author} {\bibfnamefont {D.}~\bibnamefont
	  {Loss}},\ }\href@noop {} {\bibfield  {journal} {\bibinfo  {journal} {Phys.
	  Rev. B}\ }\textbf {\bibinfo {volume} {63}},\ \bibinfo {pages} {165314}
	  (\bibinfo {year} {2001})}\BibitemShut {NoStop}%
	\bibitem [{\citenamefont {Sauret}\ \emph {et~al.}(2004)\citenamefont {Sauret},
	  \citenamefont {Feinberg},\ and\ \citenamefont {Martin}}]{SauretPRB2004a}%
	  \BibitemOpen
	  \bibfield  {author} {\bibinfo {author} {\bibfnamefont {O.}~\bibnamefont
	  {Sauret}}, \bibinfo {author} {\bibfnamefont {D.}~\bibnamefont {Feinberg}}, \
	  and\ \bibinfo {author} {\bibfnamefont {T.}~\bibnamefont {Martin}},\
	  }\href@noop {} {\bibfield  {journal} {\bibinfo  {journal} {Phys. Rev. B}\
	  }\textbf {\bibinfo {volume} {70}},\ \bibinfo {pages} {245313} (\bibinfo
	  {year} {2004})}\BibitemShut {NoStop}%
	\bibitem [{\citenamefont {Hofstetter}\ \emph {et~al.}(2009)\citenamefont
	  {Hofstetter}, \citenamefont {Csonka}, \citenamefont {Nygard},\ and\
	  \citenamefont {Sch\"onenberger}}]{HofstetterNature2009a}%
	  \BibitemOpen
	  \bibfield  {author} {\bibinfo {author} {\bibfnamefont {L.}~\bibnamefont
	  {Hofstetter}}, \bibinfo {author} {\bibfnamefont {S.}~\bibnamefont {Csonka}},
	  \bibinfo {author} {\bibfnamefont {J.}~\bibnamefont {Nygard}}, \ and\ \bibinfo
	  {author} {\bibfnamefont {C.}~\bibnamefont {Sch\"onenberger}},\ }\href@noop {}
	  {\bibfield  {journal} {\bibinfo  {journal} {Nature}\ }\textbf {\bibinfo
	  {volume} {461}},\ \bibinfo {pages} {960} (\bibinfo {year}
	  {2009})}\BibitemShut {NoStop}%
	\bibitem [{\citenamefont {Herrmann}\ \emph {et~al.}(2010)\citenamefont
	  {Herrmann}, \citenamefont {Portier}, \citenamefont {Roche}, \citenamefont
	  {Yeyati}, \citenamefont {Kontos},\ and\ \citenamefont
	  {Strunk}}]{HerrmannPRL2010a}%
	  \BibitemOpen
	  \bibfield  {author} {\bibinfo {author} {\bibfnamefont {L.~G.}\ \bibnamefont
	  {Herrmann}}, \bibinfo {author} {\bibfnamefont {F.}~\bibnamefont {Portier}},
	  \bibinfo {author} {\bibfnamefont {P.}~\bibnamefont {Roche}}, \bibinfo
	  {author} {\bibfnamefont {A.~L.}\ \bibnamefont {Yeyati}}, \bibinfo {author}
	  {\bibfnamefont {T.}~\bibnamefont {Kontos}}, \ and\ \bibinfo {author}
	  {\bibfnamefont {C.}~\bibnamefont {Strunk}},\ }\href@noop {} {\bibfield
	  {journal} {\bibinfo  {journal} {Phys. Rev. Lett.}\ }\textbf {\bibinfo
	  {volume} {104}},\ \bibinfo {pages} {026801} (\bibinfo {year}
	  {2010})}\BibitemShut {NoStop}%
	\bibitem [{\citenamefont {Schindele}\ \emph {et~al.}(2014)\citenamefont
	  {Schindele}, \citenamefont {Baumgartner}, \citenamefont {Maurand},
	  \citenamefont {Weiss},\ and\ \citenamefont
	  {Sch\"onenberger}}]{SchindelePRB2014a}%
	  \BibitemOpen
	  \bibfield  {author} {\bibinfo {author} {\bibfnamefont {J.}~\bibnamefont
	  {Schindele}}, \bibinfo {author} {\bibfnamefont {A.}~\bibnamefont
	  {Baumgartner}}, \bibinfo {author} {\bibfnamefont {R.}~\bibnamefont
	  {Maurand}}, \bibinfo {author} {\bibfnamefont {M.}~\bibnamefont {Weiss}}, \
	  and\ \bibinfo {author} {\bibfnamefont {C.}~\bibnamefont {Sch\"onenberger}},\
	  }\href@noop {} {\bibfield  {journal} {\bibinfo  {journal} {Phys. Rev. B}\
	  }\textbf {\bibinfo {volume} {89}},\ \bibinfo {pages} {045422} (\bibinfo
	  {year} {2014})}\BibitemShut {NoStop}%
	\bibitem [{\citenamefont {F\"ul\"op}\ \emph {et~al.}(2015)\citenamefont
	  {F\"ul\"op}, \citenamefont {Dom\'{\i}nguez}, \citenamefont {d'Hollosy},
	  \citenamefont {Baumgartner}, \citenamefont {Makk}, \citenamefont {Madsen},
	  \citenamefont {Guzenko}, \citenamefont {Nyg\aa{}rd}, \citenamefont
	  {Sch\"onenberger}, \citenamefont {Levy~Yeyati},\ and\ \citenamefont
	  {Csonka}}]{FueloepPRL2015a}%
	  \BibitemOpen
	  \bibfield  {author} {\bibinfo {author} {\bibfnamefont {G.}~\bibnamefont
	  {F\"ul\"op}}, \bibinfo {author} {\bibfnamefont {F.}~\bibnamefont
	  {Dom\'{\i}nguez}}, \bibinfo {author} {\bibfnamefont {S.}~\bibnamefont
	  {d'Hollosy}}, \bibinfo {author} {\bibfnamefont {A.}~\bibnamefont
	  {Baumgartner}}, \bibinfo {author} {\bibfnamefont {P.}~\bibnamefont {Makk}},
	  \bibinfo {author} {\bibfnamefont {M.~H.}\ \bibnamefont {Madsen}}, \bibinfo
	  {author} {\bibfnamefont {V.~A.}\ \bibnamefont {Guzenko}}, \bibinfo {author}
	  {\bibfnamefont {J.}~\bibnamefont {Nyg\aa{}rd}}, \bibinfo {author}
	  {\bibfnamefont {C.}~\bibnamefont {Sch\"onenberger}}, \bibinfo {author}
	  {\bibfnamefont {A.}~\bibnamefont {Levy~Yeyati}}, \ and\ \bibinfo {author}
	  {\bibfnamefont {S.}~\bibnamefont {Csonka}},\ }\href@noop {} {\bibfield
	  {journal} {\bibinfo  {journal} {Phys. Rev. Lett.}\ }\textbf {\bibinfo
	  {volume} {115}},\ \bibinfo {pages} {227003} (\bibinfo {year}
	  {2015})}\BibitemShut {NoStop}%
	\bibitem [{\citenamefont {Probst}\ \emph {et~al.}(2016)\citenamefont {Probst},
	  \citenamefont {Dom\'{\i}nguez}, \citenamefont {Schroer}, \citenamefont
	  {Yeyati},\ and\ \citenamefont {Recher}}]{ProbstPRB2016a}%
	  \BibitemOpen
	  \bibfield  {author} {\bibinfo {author} {\bibfnamefont {B.}~\bibnamefont
	  {Probst}}, \bibinfo {author} {\bibfnamefont {F.}~\bibnamefont
	  {Dom\'{\i}nguez}}, \bibinfo {author} {\bibfnamefont {A.}~\bibnamefont
	  {Schroer}}, \bibinfo {author} {\bibfnamefont {A.~L.}\ \bibnamefont {Yeyati}},
	  \ and\ \bibinfo {author} {\bibfnamefont {P.}~\bibnamefont {Recher}},\
	  }\href@noop {} {\bibfield  {journal} {\bibinfo  {journal} {Phys. Rev. B}\
	  }\textbf {\bibinfo {volume} {94}},\ \bibinfo {pages} {155445} (\bibinfo
	  {year} {2016})}\BibitemShut {NoStop}%
	\bibitem [{\citenamefont {Hussein}\ \emph {et~al.}(2016)\citenamefont
	  {Hussein}, \citenamefont {Jaurigue}, \citenamefont {Governale},\ and\
	  \citenamefont {Braggio}}]{HusseinPRB2016a}%
	  \BibitemOpen
	  \bibfield  {author} {\bibinfo {author} {\bibfnamefont {R.}~\bibnamefont
	  {Hussein}}, \bibinfo {author} {\bibfnamefont {L.}~\bibnamefont {Jaurigue}},
	  \bibinfo {author} {\bibfnamefont {M.}~\bibnamefont {Governale}}, \ and\
	  \bibinfo {author} {\bibfnamefont {A.}~\bibnamefont {Braggio}},\ }\href@noop
	  {} {\bibfield  {journal} {\bibinfo  {journal} {Phys. Rev. B}\ }\textbf
	  {\bibinfo {volume} {94}},\ \bibinfo {pages} {235134} (\bibinfo {year}
	  {2016})}\BibitemShut {NoStop}%
	\bibitem [{\citenamefont {Hussein}\ \emph {et~al.}(2017)\citenamefont
	  {Hussein}, \citenamefont {Braggio},\ and\ \citenamefont
	  {Governale}}]{HusseinPSSB2017a}%
	  \BibitemOpen
	  \bibfield  {author} {\bibinfo {author} {\bibfnamefont {R.}~\bibnamefont
	  {Hussein}}, \bibinfo {author} {\bibfnamefont {A.}~\bibnamefont {Braggio}}, \
	  and\ \bibinfo {author} {\bibfnamefont {M.}~\bibnamefont {Governale}},\
	  }\href@noop {} {\bibfield  {journal} {\bibinfo  {journal} {Phys. Status
	  Solidi B}\ }\textbf {\bibinfo {volume} {254}},\ \bibinfo {pages} {1600603}
	  (\bibinfo {year} {2017})}\BibitemShut {NoStop}%
	\bibitem [{\citenamefont {Radousky}\ and\ \citenamefont
	  {Liang}(2012)}]{RadouskyNanotechnology2012a}%
	  \BibitemOpen
	  \bibfield  {author} {\bibinfo {author} {\bibfnamefont {H.~B.}\ \bibnamefont
	  {Radousky}}\ and\ \bibinfo {author} {\bibfnamefont {H.}~\bibnamefont
	  {Liang}},\ }\href@noop {} {\bibfield  {journal} {\bibinfo  {journal}
	  {Nanotechnology}\ }\textbf {\bibinfo {volume} {23}},\ \bibinfo {pages}
	  {502001} (\bibinfo {year} {2012})}\BibitemShut {NoStop}%
	\bibitem [{\citenamefont {Roche}\ \emph {et~al.}(2015)\citenamefont {Roche},
	  \citenamefont {Roulleau}, \citenamefont {Jullien}, \citenamefont {Jompol},
	  \citenamefont {Farrer}, \citenamefont {Ritchie},\ and\ \citenamefont
	  {Glattli}}]{RocheNatureC2015a}%
	  \BibitemOpen
	  \bibfield  {author} {\bibinfo {author} {\bibfnamefont {B.}~\bibnamefont
	  {Roche}}, \bibinfo {author} {\bibfnamefont {P.}~\bibnamefont {Roulleau}},
	  \bibinfo {author} {\bibfnamefont {T.}~\bibnamefont {Jullien}}, \bibinfo
	  {author} {\bibfnamefont {Y.}~\bibnamefont {Jompol}}, \bibinfo {author}
	  {\bibfnamefont {I.}~\bibnamefont {Farrer}}, \bibinfo {author} {\bibfnamefont
	  {D.}~\bibnamefont {Ritchie}}, \ and\ \bibinfo {author} {\bibfnamefont
	  {D.}~\bibnamefont {Glattli}},\ }\href@noop {} {\bibfield  {journal} {\bibinfo
	   {journal} {Nat. Commun.}\ }\textbf {\bibinfo {volume} {6}},\ \bibinfo
	  {pages} {6738} (\bibinfo {year} {2015})}\BibitemShut {NoStop}%
	\bibitem [{\citenamefont {Sothmann}\ \emph {et~al.}(2015)\citenamefont
	  {Sothmann}, \citenamefont {S\'anchez},\ and\ \citenamefont
	  {Jordan}}]{SothmannNanotechnology2015a}%
	  \BibitemOpen
	  \bibfield  {author} {\bibinfo {author} {\bibfnamefont {B.}~\bibnamefont
	  {Sothmann}}, \bibinfo {author} {\bibfnamefont {R.}~\bibnamefont {S\'anchez}},
	  \ and\ \bibinfo {author} {\bibfnamefont {A.~N.}\ \bibnamefont {Jordan}},\
	  }\href@noop {} {\bibfield  {journal} {\bibinfo  {journal} {Nanotechnology}\
	  }\textbf {\bibinfo {volume} {26}},\ \bibinfo {pages} {032001} (\bibinfo
	  {year} {2015})}\BibitemShut {NoStop}%
	\bibitem [{\citenamefont {Thierschmann}\ \emph {et~al.}(2015)\citenamefont
	  {Thierschmann}, \citenamefont {S\'anchez}, \citenamefont {Sothmann},
	  \citenamefont {Arnold}, \citenamefont {Heyn}, \citenamefont {Hansen},
	  \citenamefont {Buhmann},\ and\ \citenamefont
	  {Molenkamp}}]{ThierschmannNatureN2015a}%
	  \BibitemOpen
	  \bibfield  {author} {\bibinfo {author} {\bibfnamefont {H.}~\bibnamefont
	  {Thierschmann}}, \bibinfo {author} {\bibfnamefont {R.}~\bibnamefont
	  {S\'anchez}}, \bibinfo {author} {\bibfnamefont {B.}~\bibnamefont {Sothmann}},
	  \bibinfo {author} {\bibfnamefont {F.}~\bibnamefont {Arnold}}, \bibinfo
	  {author} {\bibfnamefont {C.}~\bibnamefont {Heyn}}, \bibinfo {author}
	  {\bibfnamefont {W.}~\bibnamefont {Hansen}}, \bibinfo {author} {\bibfnamefont
	  {H.}~\bibnamefont {Buhmann}}, \ and\ \bibinfo {author} {\bibfnamefont
	  {L.~W.}\ \bibnamefont {Molenkamp}},\ }\href@noop {} {\bibfield  {journal}
	  {\bibinfo  {journal} {Nat. Nanotechnol.}\ }\textbf {\bibinfo {volume} {10}},\
	  \bibinfo {pages} {854} (\bibinfo {year} {2015})}\BibitemShut {NoStop}%
	\bibitem [{\citenamefont {Mastom{\"a}ki}\ \emph {et~al.}(2017)\citenamefont
	  {Mastom{\"a}ki}, \citenamefont {Roddaro}, \citenamefont {Rocci},
	  \citenamefont {Zannier}, \citenamefont {Ercolani}, \citenamefont {Sorba},
	  \citenamefont {Maasilta}, \citenamefont {Ligato}, \citenamefont {Fornieri},
	  \citenamefont {Strambini},\ and\ \citenamefont
	  {Giazotto}}]{MastomaekiNanoR2017a}%
	  \BibitemOpen
	  \bibfield  {author} {\bibinfo {author} {\bibfnamefont {J.}~\bibnamefont
	  {Mastom{\"a}ki}}, \bibinfo {author} {\bibfnamefont {S.}~\bibnamefont
	  {Roddaro}}, \bibinfo {author} {\bibfnamefont {M.}~\bibnamefont {Rocci}},
	  \bibinfo {author} {\bibfnamefont {V.}~\bibnamefont {Zannier}}, \bibinfo
	  {author} {\bibfnamefont {D.}~\bibnamefont {Ercolani}}, \bibinfo {author}
	  {\bibfnamefont {L.}~\bibnamefont {Sorba}}, \bibinfo {author} {\bibfnamefont
	  {I.~J.}\ \bibnamefont {Maasilta}}, \bibinfo {author} {\bibfnamefont
	  {N.}~\bibnamefont {Ligato}}, \bibinfo {author} {\bibfnamefont
	  {A.}~\bibnamefont {Fornieri}}, \bibinfo {author} {\bibfnamefont
	  {E.}~\bibnamefont {Strambini}}, \ and\ \bibinfo {author} {\bibfnamefont
	  {F.}~\bibnamefont {Giazotto}},\ }\href@noop {} {\bibfield  {journal}
	  {\bibinfo  {journal} {Nano Res.}\ }\textbf {\bibinfo {volume} {10}},\
	  \bibinfo {pages} {3468} (\bibinfo {year} {2017})}\BibitemShut {NoStop}%
	\bibitem [{\citenamefont {Virtanen}\ and\ \citenamefont
	  {Heikkil\"a}(2004)}]{VirtanenPRL2004a}%
	  \BibitemOpen
	  \bibfield  {author} {\bibinfo {author} {\bibfnamefont {P.}~\bibnamefont
	  {Virtanen}}\ and\ \bibinfo {author} {\bibfnamefont {T.~T.}\ \bibnamefont
	  {Heikkil\"a}},\ }\href@noop {} {\bibfield  {journal} {\bibinfo  {journal}
	  {Phys. Rev. Lett.}\ }\textbf {\bibinfo {volume} {92}},\ \bibinfo {pages}
	  {177004} (\bibinfo {year} {2004})}\BibitemShut {NoStop}%
	\bibitem [{\citenamefont {Machon}\ \emph {et~al.}(2013)\citenamefont {Machon},
	  \citenamefont {Eschrig},\ and\ \citenamefont {Belzig}}]{MachonPRL2013a}%
	  \BibitemOpen
	  \bibfield  {author} {\bibinfo {author} {\bibfnamefont {P.}~\bibnamefont
	  {Machon}}, \bibinfo {author} {\bibfnamefont {M.}~\bibnamefont {Eschrig}}, \
	  and\ \bibinfo {author} {\bibfnamefont {W.}~\bibnamefont {Belzig}},\
	  }\href@noop {} {\bibfield  {journal} {\bibinfo  {journal} {Phys. Rev. Lett.}\
	  }\textbf {\bibinfo {volume} {110}},\ \bibinfo {pages} {047002} (\bibinfo
	  {year} {2013})}\BibitemShut {NoStop}%
	\bibitem [{\citenamefont {Machon}\ \emph {et~al.}(2014)\citenamefont {Machon},
	  \citenamefont {Eschrig},\ and\ \citenamefont {Belzig}}]{MachonNJP2014a}%
	  \BibitemOpen
	  \bibfield  {author} {\bibinfo {author} {\bibfnamefont {P.}~\bibnamefont
	  {Machon}}, \bibinfo {author} {\bibfnamefont {M.}~\bibnamefont {Eschrig}}, \
	  and\ \bibinfo {author} {\bibfnamefont {W.}~\bibnamefont {Belzig}},\
	  }\href@noop {} {\bibfield  {journal} {\bibinfo  {journal} {New J. Phys.}\
	  }\textbf {\bibinfo {volume} {16}},\ \bibinfo {pages} {073002} (\bibinfo
	  {year} {2014})}\BibitemShut {NoStop}%
	\bibitem [{\citenamefont {Ozaeta}\ \emph {et~al.}(2014)\citenamefont {Ozaeta},
	  \citenamefont {Virtanen}, \citenamefont {Bergeret},\ and\ \citenamefont
	  {Heikkil\"a}}]{OzaetaPRL2014a}%
	  \BibitemOpen
	  \bibfield  {author} {\bibinfo {author} {\bibfnamefont {A.}~\bibnamefont
	  {Ozaeta}}, \bibinfo {author} {\bibfnamefont {P.}~\bibnamefont {Virtanen}},
	  \bibinfo {author} {\bibfnamefont {F.~S.}\ \bibnamefont {Bergeret}}, \ and\
	  \bibinfo {author} {\bibfnamefont {T.~T.}\ \bibnamefont {Heikkil\"a}},\
	  }\href@noop {} {\bibfield  {journal} {\bibinfo  {journal} {Phys. Rev. Lett.}\
	  }\textbf {\bibinfo {volume} {112}},\ \bibinfo {pages} {057001} (\bibinfo
	  {year} {2014})}\BibitemShut {NoStop}%
	\bibitem [{\citenamefont {Giazotto}\ \emph
	  {et~al.}(2015{\natexlab{a}})\citenamefont {Giazotto}, \citenamefont
	  {Solinas}, \citenamefont {Braggio},\ and\ \citenamefont
	  {Bergeret}}]{GiazottoPRApplied2015a}%
	  \BibitemOpen
	  \bibfield  {author} {\bibinfo {author} {\bibfnamefont {F.}~\bibnamefont
	  {Giazotto}}, \bibinfo {author} {\bibfnamefont {P.}~\bibnamefont {Solinas}},
	  \bibinfo {author} {\bibfnamefont {A.}~\bibnamefont {Braggio}}, \ and\
	  \bibinfo {author} {\bibfnamefont {F.~S.}\ \bibnamefont {Bergeret}},\
	  }\href@noop {} {\bibfield  {journal} {\bibinfo  {journal} {Phys. Rev.
	  Applied}\ }\textbf {\bibinfo {volume} {4}},\ \bibinfo {pages} {044016}
	  (\bibinfo {year} {2015}{\natexlab{a}})}\BibitemShut {NoStop}%
	\bibitem [{\citenamefont {Giazotto}\ \emph
	  {et~al.}(2015{\natexlab{b}})\citenamefont {Giazotto}, \citenamefont
	  {Heikkil\"a},\ and\ \citenamefont {Bergeret}}]{GiazottoPRL2015a}%
	  \BibitemOpen
	  \bibfield  {author} {\bibinfo {author} {\bibfnamefont {F.}~\bibnamefont
	  {Giazotto}}, \bibinfo {author} {\bibfnamefont {T.~T.}\ \bibnamefont
	  {Heikkil\"a}}, \ and\ \bibinfo {author} {\bibfnamefont {F.~S.}\ \bibnamefont
	  {Bergeret}},\ }\href@noop {} {\bibfield  {journal} {\bibinfo  {journal}
	  {Phys. Rev. Lett.}\ }\textbf {\bibinfo {volume} {114}},\ \bibinfo {pages}
	  {067001} (\bibinfo {year} {2015}{\natexlab{b}})}\BibitemShut {NoStop}%
	\bibitem [{\citenamefont {Linder}\ and\ \citenamefont
	  {Bathen}(2016)}]{LinderPRB2016a}%
	  \BibitemOpen
	  \bibfield  {author} {\bibinfo {author} {\bibfnamefont {J.}~\bibnamefont
	  {Linder}}\ and\ \bibinfo {author} {\bibfnamefont {M.~E.}\ \bibnamefont
	  {Bathen}},\ }\href@noop {} {\bibfield  {journal} {\bibinfo  {journal} {Phys.
	  Rev. B}\ }\textbf {\bibinfo {volume} {93}},\ \bibinfo {pages} {224509}
	  (\bibinfo {year} {2016})}\BibitemShut {NoStop}%
	\bibitem [{\citenamefont {Wysoki\'nski}(2012)}]{WysokinskiJP2012a}%
	  \BibitemOpen
	  \bibfield  {author} {\bibinfo {author} {\bibfnamefont {K.~I.}\ \bibnamefont
	  {Wysoki\'nski}},\ }\href@noop {} {\bibfield  {journal} {\bibinfo  {journal}
	  {J. Phys.: Condens. Matter}\ }\textbf {\bibinfo {volume} {24}},\ \bibinfo
	  {pages} {335303} (\bibinfo {year} {2012})}\BibitemShut {NoStop}%
	\bibitem [{\citenamefont {Hwang}\ \emph {et~al.}(2015)\citenamefont {Hwang},
	  \citenamefont {L\'opez},\ and\ \citenamefont {S\'anchez}}]{HwangPRB2015a}%
	  \BibitemOpen
	  \bibfield  {author} {\bibinfo {author} {\bibfnamefont {S.-Y.}\ \bibnamefont
	  {Hwang}}, \bibinfo {author} {\bibfnamefont {R.}~\bibnamefont {L\'opez}}, \
	  and\ \bibinfo {author} {\bibfnamefont {D.}~\bibnamefont {S\'anchez}},\
	  }\href@noop {} {\bibfield  {journal} {\bibinfo  {journal} {Phys. Rev. B}\
	  }\textbf {\bibinfo {volume} {91}},\ \bibinfo {pages} {104518} (\bibinfo
	  {year} {2015})}\BibitemShut {NoStop}%
	\bibitem [{\citenamefont {Hwang}\ \emph {et~al.}(2016)\citenamefont {Hwang},
	  \citenamefont {L\'opez},\ and\ \citenamefont {S\'anchez}}]{HwangPRB2016a}%
	  \BibitemOpen
	  \bibfield  {author} {\bibinfo {author} {\bibfnamefont {S.-Y.}\ \bibnamefont
	  {Hwang}}, \bibinfo {author} {\bibfnamefont {R.}~\bibnamefont {L\'opez}}, \
	  and\ \bibinfo {author} {\bibfnamefont {D.}~\bibnamefont {S\'anchez}},\
	  }\href@noop {} {\bibfield  {journal} {\bibinfo  {journal} {Phys. Rev. B}\
	  }\textbf {\bibinfo {volume} {94}},\ \bibinfo {pages} {054506} (\bibinfo
	  {year} {2016})}\BibitemShut {NoStop}%
	\bibitem [{\citenamefont {Trocha}\ and\ \citenamefont
	  {Barna\ifmmode~\acute{s}\else \'{s}\fi{}}(2017)}]{TrochaPRB2017a}%
	  \BibitemOpen
	  \bibfield  {author} {\bibinfo {author} {\bibfnamefont {P.}~\bibnamefont
	  {Trocha}}\ and\ \bibinfo {author} {\bibfnamefont {J.}~\bibnamefont
	  {Barna\ifmmode~\acute{s}\else \'{s}\fi{}}},\ }\href@noop {} {\bibfield
	  {journal} {\bibinfo  {journal} {Phys. Rev. B}\ }\textbf {\bibinfo {volume}
	  {95}},\ \bibinfo {pages} {165439} (\bibinfo {year} {2017})}\BibitemShut
	  {NoStop}%
	\bibitem [{\citenamefont {Van~den Broeck}(2005)}]{VandenBroeckPRL2005a}%
	  \BibitemOpen
	  \bibfield  {author} {\bibinfo {author} {\bibfnamefont {C.}~\bibnamefont
	  {Van~den Broeck}},\ }\href@noop {} {\bibfield  {journal} {\bibinfo  {journal}
	  {Phys. Rev. Lett.}\ }\textbf {\bibinfo {volume} {95}},\ \bibinfo {pages}
	  {190602} (\bibinfo {year} {2005})}\BibitemShut {NoStop}%
	\bibitem [{\citenamefont {Muralidharan}\ and\ \citenamefont
	  {Grifoni}(2012)}]{MuralidharanPRB2012a}%
	  \BibitemOpen
	  \bibfield  {author} {\bibinfo {author} {\bibfnamefont {B.}~\bibnamefont
	  {Muralidharan}}\ and\ \bibinfo {author} {\bibfnamefont {M.}~\bibnamefont
	  {Grifoni}},\ }\href@noop {} {\bibfield  {journal} {\bibinfo  {journal} {Phys.
	  Rev. B}\ }\textbf {\bibinfo {volume} {85}},\ \bibinfo {pages} {155423}
	  (\bibinfo {year} {2012})}\BibitemShut {NoStop}%
	\bibitem [{\citenamefont {Brunner}\ \emph {et~al.}(2012)\citenamefont
	  {Brunner}, \citenamefont {Linden}, \citenamefont {Popescu},\ and\
	  \citenamefont {Skrzypczyk}}]{BrunnerPRE2012a}%
	  \BibitemOpen
	  \bibfield  {author} {\bibinfo {author} {\bibfnamefont {N.}~\bibnamefont
	  {Brunner}}, \bibinfo {author} {\bibfnamefont {N.}~\bibnamefont {Linden}},
	  \bibinfo {author} {\bibfnamefont {S.}~\bibnamefont {Popescu}}, \ and\
	  \bibinfo {author} {\bibfnamefont {P.}~\bibnamefont {Skrzypczyk}},\
	  }\href@noop {} {\bibfield  {journal} {\bibinfo  {journal} {Phys. Rev. E}\
	  }\textbf {\bibinfo {volume} {85}},\ \bibinfo {pages} {051117} (\bibinfo
	  {year} {2012})}\BibitemShut {NoStop}%
	\bibitem [{\citenamefont {Correa}\ \emph {et~al.}(2013)\citenamefont {Correa},
	  \citenamefont {Palao}, \citenamefont {Adesso},\ and\ \citenamefont
	  {Alonso}}]{CorreaPRE2013a}%
	  \BibitemOpen
	  \bibfield  {author} {\bibinfo {author} {\bibfnamefont {L.~A.}\ \bibnamefont
	  {Correa}}, \bibinfo {author} {\bibfnamefont {J.~P.}\ \bibnamefont {Palao}},
	  \bibinfo {author} {\bibfnamefont {G.}~\bibnamefont {Adesso}}, \ and\ \bibinfo
	  {author} {\bibfnamefont {D.}~\bibnamefont {Alonso}},\ }\href@noop {}
	  {\bibfield  {journal} {\bibinfo  {journal} {Phys. Rev. E}\ }\textbf {\bibinfo
	  {volume} {87}},\ \bibinfo {pages} {042131} (\bibinfo {year}
	  {2013})}\BibitemShut {NoStop}%
	\bibitem [{\citenamefont {Mazza}\ \emph {et~al.}(2014)\citenamefont {Mazza},
	  \citenamefont {Bosisio}, \citenamefont {Benenti}, \citenamefont
	  {Giovannetti}, \citenamefont {Fazio},\ and\ \citenamefont
	  {Taddei}}]{MazzaNJP2014a}%
	  \BibitemOpen
	  \bibfield  {author} {\bibinfo {author} {\bibfnamefont {F.}~\bibnamefont
	  {Mazza}}, \bibinfo {author} {\bibfnamefont {R.}~\bibnamefont {Bosisio}},
	  \bibinfo {author} {\bibfnamefont {G.}~\bibnamefont {Benenti}}, \bibinfo
	  {author} {\bibfnamefont {V.}~\bibnamefont {Giovannetti}}, \bibinfo {author}
	  {\bibfnamefont {R.}~\bibnamefont {Fazio}}, \ and\ \bibinfo {author}
	  {\bibfnamefont {F.}~\bibnamefont {Taddei}},\ }\href@noop {} {\bibfield
	  {journal} {\bibinfo  {journal} {New J. Phys.}\ }\textbf {\bibinfo {volume}
	  {16}},\ \bibinfo {pages} {085001} (\bibinfo {year} {2014})}\BibitemShut
	  {NoStop}%
	\bibitem [{\citenamefont {Whitney}(2014)}]{WhitneyPRL2014a}%
	  \BibitemOpen
	  \bibfield  {author} {\bibinfo {author} {\bibfnamefont {R.~S.}\ \bibnamefont
	  {Whitney}},\ }\href@noop {} {\bibfield  {journal} {\bibinfo  {journal} {Phys.
	  Rev. Lett.}\ }\textbf {\bibinfo {volume} {112}},\ \bibinfo {pages} {130601}
	  (\bibinfo {year} {2014})}\BibitemShut {NoStop}%
	\bibitem [{\citenamefont {Hofer}\ \emph {et~al.}(2016)\citenamefont {Hofer},
	  \citenamefont {Souquet},\ and\ \citenamefont {Clerk}}]{HoferPRB2016a}%
	  \BibitemOpen
	  \bibfield  {author} {\bibinfo {author} {\bibfnamefont {P.~P.}\ \bibnamefont
	  {Hofer}}, \bibinfo {author} {\bibfnamefont {J.-R.}\ \bibnamefont {Souquet}},
	  \ and\ \bibinfo {author} {\bibfnamefont {A.~A.}\ \bibnamefont {Clerk}},\
	  }\href@noop {} {\bibfield  {journal} {\bibinfo  {journal} {Phys. Rev. B}\
	  }\textbf {\bibinfo {volume} {93}},\ \bibinfo {pages} {041418} (\bibinfo
	  {year} {2016})}\BibitemShut {NoStop}%
	\bibitem [{\citenamefont {Benenti}\ \emph {et~al.}(2017)\citenamefont
	  {Benenti}, \citenamefont {Casati}, \citenamefont {Saito},\ and\ \citenamefont
	  {Whitney}}]{BenentiPR2017a}%
	  \BibitemOpen
	  \bibfield  {author} {\bibinfo {author} {\bibfnamefont {G.}~\bibnamefont
	  {Benenti}}, \bibinfo {author} {\bibfnamefont {G.}~\bibnamefont {Casati}},
	  \bibinfo {author} {\bibfnamefont {K.}~\bibnamefont {Saito}}, \ and\ \bibinfo
	  {author} {\bibfnamefont {R.~S.}\ \bibnamefont {Whitney}},\ }\href@noop {}
	  {\bibfield  {journal} {\bibinfo  {journal} {Phys. Rep.}\ }\textbf {\bibinfo
	  {volume} {694}},\ \bibinfo {pages} {1} (\bibinfo {year} {2017})}\BibitemShut
	  {NoStop}%
	\bibitem [{\citenamefont {Seah}\ \emph {et~al.}(2018)\citenamefont {Seah},
	  \citenamefont {Nimmrichter},\ and\ \citenamefont {Scarani}}]{SeahPRE2018a}%
	  \BibitemOpen
	  \bibfield  {author} {\bibinfo {author} {\bibfnamefont {S.}~\bibnamefont
	  {Seah}}, \bibinfo {author} {\bibfnamefont {S.}~\bibnamefont {Nimmrichter}}, \
	  and\ \bibinfo {author} {\bibfnamefont {V.}~\bibnamefont {Scarani}},\
	  }\href@noop {} {\bibfield  {journal} {\bibinfo  {journal} {Phys. Rev. E}\
	  }\textbf {\bibinfo {volume} {98}},\ \bibinfo {pages} {012131} (\bibinfo
	  {year} {2018})}\BibitemShut {NoStop}%
	\bibitem [{\citenamefont {Karwacki}\ and\ \citenamefont
	  {Trocha}(2016)}]{KarwackiPRB2016a}%
	  \BibitemOpen
	  \bibfield  {author} {\bibinfo {author} {\bibfnamefont {L.}~\bibnamefont
	  {Karwacki}}\ and\ \bibinfo {author} {\bibfnamefont {P.}~\bibnamefont
	  {Trocha}},\ }\href@noop {} {\bibfield  {journal} {\bibinfo  {journal} {Phys.
	  Rev. B}\ }\textbf {\bibinfo {volume} {94}},\ \bibinfo {pages} {085418}
	  (\bibinfo {year} {2016})}\BibitemShut {NoStop}%
	\bibitem [{\citenamefont {Erdman}\ \emph {et~al.}(2017)\citenamefont {Erdman},
	  \citenamefont {Mazza}, \citenamefont {Bosisio}, \citenamefont {Benenti},
	  \citenamefont {Fazio},\ and\ \citenamefont {Taddei}}]{ErdmanPRB2017a}%
	  \BibitemOpen
	  \bibfield  {author} {\bibinfo {author} {\bibfnamefont {P.~A.}\ \bibnamefont
	  {Erdman}}, \bibinfo {author} {\bibfnamefont {F.}~\bibnamefont {Mazza}},
	  \bibinfo {author} {\bibfnamefont {R.}~\bibnamefont {Bosisio}}, \bibinfo
	  {author} {\bibfnamefont {G.}~\bibnamefont {Benenti}}, \bibinfo {author}
	  {\bibfnamefont {R.}~\bibnamefont {Fazio}}, \ and\ \bibinfo {author}
	  {\bibfnamefont {F.}~\bibnamefont {Taddei}},\ }\href@noop {} {\bibfield
	  {journal} {\bibinfo  {journal} {Phys. Rev. B}\ }\textbf {\bibinfo {volume}
	  {95}},\ \bibinfo {pages} {245432} (\bibinfo {year} {2017})}\BibitemShut
	  {NoStop}%
	\bibitem [{\citenamefont {Cao}\ \emph {et~al.}(2015)\citenamefont {Cao},
	  \citenamefont {Fang}, \citenamefont {Li},\ and\ \citenamefont
	  {Luo}}]{CaoAPL2015a}%
	  \BibitemOpen
	  \bibfield  {author} {\bibinfo {author} {\bibfnamefont {Z.}~\bibnamefont
	  {Cao}}, \bibinfo {author} {\bibfnamefont {T.-F.}\ \bibnamefont {Fang}},
	  \bibinfo {author} {\bibfnamefont {L.}~\bibnamefont {Li}}, \ and\ \bibinfo
	  {author} {\bibfnamefont {H.-G.}\ \bibnamefont {Luo}},\ }\href@noop {}
	  {\bibfield  {journal} {\bibinfo  {journal} {Appl. Phys. Lett.}\ }\textbf
	  {\bibinfo {volume} {107}},\ \bibinfo {eid} {212601} (\bibinfo {year}
	  {2015})}\BibitemShut {NoStop}%
	\bibitem [{\citenamefont {Rozhkov}\ and\ \citenamefont
	  {Arovas}(2000)}]{RozhkovPRB2000a}%
	  \BibitemOpen
	  \bibfield  {author} {\bibinfo {author} {\bibfnamefont {A.~V.}\ \bibnamefont
	  {Rozhkov}}\ and\ \bibinfo {author} {\bibfnamefont {D.~P.}\ \bibnamefont
	  {Arovas}},\ }\href@noop {} {\bibfield  {journal} {\bibinfo  {journal} {Phys.
	  Rev. B}\ }\textbf {\bibinfo {volume} {62}},\ \bibinfo {pages} {6687}
	  (\bibinfo {year} {2000})}\BibitemShut {NoStop}%
	\bibitem [{\citenamefont {Meng}\ \emph {et~al.}(2009)\citenamefont {Meng},
	  \citenamefont {Florens},\ and\ \citenamefont {Simon}}]{MengPRB2009a}%
	  \BibitemOpen
	  \bibfield  {author} {\bibinfo {author} {\bibfnamefont {T.}~\bibnamefont
	  {Meng}}, \bibinfo {author} {\bibfnamefont {S.}~\bibnamefont {Florens}}, \
	  and\ \bibinfo {author} {\bibfnamefont {P.}~\bibnamefont {Simon}},\
	  }\href@noop {} {\bibfield  {journal} {\bibinfo  {journal} {Phys. Rev. B}\
	  }\textbf {\bibinfo {volume} {79}},\ \bibinfo {pages} {224521} (\bibinfo
	  {year} {2009})}\BibitemShut {NoStop}%
	\bibitem [{\citenamefont {Eldridge}\ \emph {et~al.}(2010)\citenamefont
	  {Eldridge}, \citenamefont {Pala}, \citenamefont {Governale},\ and\
	  \citenamefont {K\"onig}}]{EldridgePRB2010a}%
	  \BibitemOpen
	  \bibfield  {author} {\bibinfo {author} {\bibfnamefont {J.}~\bibnamefont
	  {Eldridge}}, \bibinfo {author} {\bibfnamefont {M.~G.}\ \bibnamefont {Pala}},
	  \bibinfo {author} {\bibfnamefont {M.}~\bibnamefont {Governale}}, \ and\
	  \bibinfo {author} {\bibfnamefont {J.}~\bibnamefont {K\"onig}},\ }\href@noop
	  {} {\bibfield  {journal} {\bibinfo  {journal} {Phys. Rev. B}\ }\textbf
	  {\bibinfo {volume} {82}},\ \bibinfo {pages} {184507} (\bibinfo {year}
	  {2010})}\BibitemShut {NoStop}%
	\bibitem [{\citenamefont {Braggio}\ \emph {et~al.}(2011)\citenamefont
	  {Braggio}, \citenamefont {Governale}, \citenamefont {Pala},\ and\
	  \citenamefont {K\"onig}}]{BraggioSSC2011a}%
	  \BibitemOpen
	  \bibfield  {author} {\bibinfo {author} {\bibfnamefont {A.}~\bibnamefont
	  {Braggio}}, \bibinfo {author} {\bibfnamefont {M.}~\bibnamefont {Governale}},
	  \bibinfo {author} {\bibfnamefont {M.~G.}\ \bibnamefont {Pala}}, \ and\
	  \bibinfo {author} {\bibfnamefont {J.}~\bibnamefont {K\"onig}},\ }\href@noop
	  {} {\bibfield  {journal} {\bibinfo  {journal} {Solid State Commun.}\ }\textbf
	  {\bibinfo {volume} {151}},\ \bibinfo {pages} {155} (\bibinfo {year}
	  {2011})}\BibitemShut {NoStop}%
	\bibitem [{\citenamefont {Sothmann}\ \emph {et~al.}(2014)\citenamefont
	  {Sothmann}, \citenamefont {Weiss}, \citenamefont {Governale},\ and\
	  \citenamefont {K\"onig}}]{SothmannPRB2014a}%
	  \BibitemOpen
	  \bibfield  {author} {\bibinfo {author} {\bibfnamefont {B.}~\bibnamefont
	  {Sothmann}}, \bibinfo {author} {\bibfnamefont {S.}~\bibnamefont {Weiss}},
	  \bibinfo {author} {\bibfnamefont {M.}~\bibnamefont {Governale}}, \ and\
	  \bibinfo {author} {\bibfnamefont {J.}~\bibnamefont {K\"onig}},\ }\href@noop
	  {} {\bibfield  {journal} {\bibinfo  {journal} {Phys. Rev. B}\ }\textbf
	  {\bibinfo {volume} {90}},\ \bibinfo {pages} {220501} (\bibinfo {year}
	  {2014})}\BibitemShut {NoStop}%
	\bibitem [{\citenamefont {Weiss}\ and\ \citenamefont
	  {K\"onig}(2017)}]{WeissPRB2017a}%
	  \BibitemOpen
	  \bibfield  {author} {\bibinfo {author} {\bibfnamefont {S.}~\bibnamefont
	  {Weiss}}\ and\ \bibinfo {author} {\bibfnamefont {J.}~\bibnamefont
	  {K\"onig}},\ }\href@noop {} {\bibfield  {journal} {\bibinfo  {journal} {Phys.
	  Rev. B}\ }\textbf {\bibinfo {volume} {96}},\ \bibinfo {pages} {064529}
	  (\bibinfo {year} {2017})}\BibitemShut {NoStop}%
	\bibitem [{\citenamefont {Walldorf}\ \emph {et~al.}(2018)\citenamefont
	  {Walldorf}, \citenamefont {Padurariu}, \citenamefont {Jauho},\ and\
	  \citenamefont {Flindt}}]{WalldorfPRL2018a}%
	  \BibitemOpen
	  \bibfield  {author} {\bibinfo {author} {\bibfnamefont {N.}~\bibnamefont
	  {Walldorf}}, \bibinfo {author} {\bibfnamefont {C.}~\bibnamefont {Padurariu}},
	  \bibinfo {author} {\bibfnamefont {A.-P.}\ \bibnamefont {Jauho}}, \ and\
	  \bibinfo {author} {\bibfnamefont {C.}~\bibnamefont {Flindt}},\ }\href@noop {}
	  {\bibfield  {journal} {\bibinfo  {journal} {Phys. Rev. Lett.}\ }\textbf
	  {\bibinfo {volume} {120}},\ \bibinfo {pages} {087701} (\bibinfo {year}
	  {2018})}\BibitemShut {NoStop}%
	\bibitem [{\citenamefont {Kaiser}\ \emph {et~al.}(2006)\citenamefont {Kaiser},
	  \citenamefont {Strass}, \citenamefont {Kohler},\ and\ \citenamefont
	  {H\"anggi}}]{KaiserCP2006a}%
	  \BibitemOpen
	  \bibfield  {author} {\bibinfo {author} {\bibfnamefont {F.~J.}\ \bibnamefont
	  {Kaiser}}, \bibinfo {author} {\bibfnamefont {M.}~\bibnamefont {Strass}},
	  \bibinfo {author} {\bibfnamefont {S.}~\bibnamefont {Kohler}}, \ and\ \bibinfo
	  {author} {\bibfnamefont {P.}~\bibnamefont {H\"anggi}},\ }\href@noop {}
	  {\bibfield  {journal} {\bibinfo  {journal} {Chem. Phys.}\ }\textbf {\bibinfo
	  {volume} {322}},\ \bibinfo {pages} {193} (\bibinfo {year}
	  {2006})}\BibitemShut {NoStop}%
	\bibitem [{\citenamefont {Darau}\ \emph {et~al.}(2009)\citenamefont {Darau},
	  \citenamefont {Begemann}, \citenamefont {Donarini},\ and\ \citenamefont
	  {Grifoni}}]{DarauPRB2009a}%
	  \BibitemOpen
	  \bibfield  {author} {\bibinfo {author} {\bibfnamefont {D.}~\bibnamefont
	  {Darau}}, \bibinfo {author} {\bibfnamefont {G.}~\bibnamefont {Begemann}},
	  \bibinfo {author} {\bibfnamefont {A.}~\bibnamefont {Donarini}}, \ and\
	  \bibinfo {author} {\bibfnamefont {M.}~\bibnamefont {Grifoni}},\ }\href@noop
	  {} {\bibfield  {journal} {\bibinfo  {journal} {Phys. Rev. B}\ }\textbf
	  {\bibinfo {volume} {79}},\ \bibinfo {pages} {235404} (\bibinfo {year}
	  {2009})}\BibitemShut {NoStop}%
	\bibitem [{\citenamefont {Hofer}\ \emph {et~al.}(2017)\citenamefont {Hofer},
	  \citenamefont {Perarnau-Llobet}, \citenamefont {Miranda}, \citenamefont
	  {Haack}, \citenamefont {Silva}, \citenamefont {Brask},\ and\ \citenamefont
	  {Brunner}}]{HoferNJP2017a}%
	  \BibitemOpen
	  \bibfield  {author} {\bibinfo {author} {\bibfnamefont {P.~P.}\ \bibnamefont
	  {Hofer}}, \bibinfo {author} {\bibfnamefont {M.}~\bibnamefont
	  {Perarnau-Llobet}}, \bibinfo {author} {\bibfnamefont {L.~D.~M.}\ \bibnamefont
	  {Miranda}}, \bibinfo {author} {\bibfnamefont {G.}~\bibnamefont {Haack}},
	  \bibinfo {author} {\bibfnamefont {R.}~\bibnamefont {Silva}}, \bibinfo
	  {author} {\bibfnamefont {J.~B.}\ \bibnamefont {Brask}}, \ and\ \bibinfo
	  {author} {\bibfnamefont {N.}~\bibnamefont {Brunner}},\ }\href@noop {}
	  {\bibfield  {journal} {\bibinfo  {journal} {New J. Phys.}\ }\textbf {\bibinfo
	  {volume} {19}},\ \bibinfo {pages} {123037} (\bibinfo {year}
	  {2017})}\BibitemShut {NoStop}%
	\bibitem [{\citenamefont {Gonz\'alez}\ \emph {et~al.}(2017)\citenamefont
	  {Gonz\'alez}, \citenamefont {Correa}, \citenamefont {Nocerino}, \citenamefont
	  {Palao}, \citenamefont {Alonso},\ and\ \citenamefont
	  {Adesso}}]{GonzalezOpenSID2017a}%
	  \BibitemOpen
	  \bibfield  {author} {\bibinfo {author} {\bibfnamefont {J.~O.}\ \bibnamefont
	  {Gonz\'alez}}, \bibinfo {author} {\bibfnamefont {L.~A.}\ \bibnamefont
	  {Correa}}, \bibinfo {author} {\bibfnamefont {G.}~\bibnamefont {Nocerino}},
	  \bibinfo {author} {\bibfnamefont {J.~P.}\ \bibnamefont {Palao}}, \bibinfo
	  {author} {\bibfnamefont {D.}~\bibnamefont {Alonso}}, \ and\ \bibinfo {author}
	  {\bibfnamefont {G.}~\bibnamefont {Adesso}},\ }\href@noop {} {\bibfield
	  {journal} {\bibinfo  {journal} {Open Syst. Inf. Dyn.}\ }\textbf {\bibinfo
	  {volume} {24}},\ \bibinfo {pages} {1740010} (\bibinfo {year}
	  {2017})}\BibitemShut {NoStop}%
	\bibitem [{\citenamefont {Mazza}\ \emph {et~al.}(2015)\citenamefont {Mazza},
	  \citenamefont {Valentini}, \citenamefont {Bosisio}, \citenamefont {Benenti},
	  \citenamefont {Giovannetti}, \citenamefont {Fazio},\ and\ \citenamefont
	  {Taddei}}]{MazzaPRB2015a}%
	  \BibitemOpen
	  \bibfield  {author} {\bibinfo {author} {\bibfnamefont {F.}~\bibnamefont
	  {Mazza}}, \bibinfo {author} {\bibfnamefont {S.}~\bibnamefont {Valentini}},
	  \bibinfo {author} {\bibfnamefont {R.}~\bibnamefont {Bosisio}}, \bibinfo
	  {author} {\bibfnamefont {G.}~\bibnamefont {Benenti}}, \bibinfo {author}
	  {\bibfnamefont {V.}~\bibnamefont {Giovannetti}}, \bibinfo {author}
	  {\bibfnamefont {R.}~\bibnamefont {Fazio}}, \ and\ \bibinfo {author}
	  {\bibfnamefont {F.}~\bibnamefont {Taddei}},\ }\href@noop {} {\bibfield
	  {journal} {\bibinfo  {journal} {Phys. Rev. B}\ }\textbf {\bibinfo {volume}
	  {91}},\ \bibinfo {pages} {245435} (\bibinfo {year} {2015})}\BibitemShut
	  {NoStop}%
	\bibitem [{Note1()}]{Note1}%
	  \BibitemOpen
	  \bibinfo {note} {Notice, that at the non-local resonance even for a finite
	  voltage bias between the normal leads essentially no current will flow
	  between them in the limit of strong intradot Coulomb
	  interaction.}\BibitemShut {Stop}%
	\bibitem [{\citenamefont {Saito}\ and\ \citenamefont
	  {Dhar}(2007)}]{SaitoPRL2007a}%
	  \BibitemOpen
	  \bibfield  {author} {\bibinfo {author} {\bibfnamefont {K.}~\bibnamefont
	  {Saito}}\ and\ \bibinfo {author} {\bibfnamefont {A.}~\bibnamefont {Dhar}},\
	  }\href@noop {} {\bibfield  {journal} {\bibinfo  {journal} {Phys. Rev. Lett.}\
	  }\textbf {\bibinfo {volume} {99}},\ \bibinfo {pages} {180601} (\bibinfo
	  {year} {2007})}\BibitemShut {NoStop}%
	\bibitem [{\citenamefont {Saito}\ and\ \citenamefont
	  {Utsumi}(2008)}]{SaitoPRB2008a}%
	  \BibitemOpen
	  \bibfield  {author} {\bibinfo {author} {\bibfnamefont {K.}~\bibnamefont
	  {Saito}}\ and\ \bibinfo {author} {\bibfnamefont {Y.}~\bibnamefont {Utsumi}},\
	  }\href@noop {} {\bibfield  {journal} {\bibinfo  {journal} {Phys. Rev. B}\
	  }\textbf {\bibinfo {volume} {78}},\ \bibinfo {pages} {115429} (\bibinfo
	  {year} {2008})}\BibitemShut {NoStop}%
	\bibitem [{\citenamefont {Hussein}\ and\ \citenamefont
	  {Kohler}(2014)}]{HusseinPRB2014a}%
	  \BibitemOpen
	  \bibfield  {author} {\bibinfo {author} {\bibfnamefont {R.}~\bibnamefont
	  {Hussein}}\ and\ \bibinfo {author} {\bibfnamefont {S.}~\bibnamefont
	  {Kohler}},\ }\href@noop {} {\bibfield  {journal} {\bibinfo  {journal} {Phys.
	  Rev. B}\ }\textbf {\bibinfo {volume} {89}},\ \bibinfo {pages} {205424}
	  (\bibinfo {year} {2014})}\BibitemShut {NoStop}%
	\bibitem [{\citenamefont {Bagrets}\ and\ \citenamefont {{Y}u.
	  V.~Nazarov}(2003)}]{BagretsPRB2003a}%
	  \BibitemOpen
	  \bibfield  {author} {\bibinfo {author} {\bibfnamefont {D.~A.}\ \bibnamefont
	  {Bagrets}}\ and\ \bibinfo {author} {\bibnamefont {{Y}u. V.~Nazarov}},\
	  }\href@noop {} {\bibfield  {journal} {\bibinfo  {journal} {Phys. Rev. B}\
	  }\textbf {\bibinfo {volume} {67}},\ \bibinfo {pages} {085316} (\bibinfo
	  {year} {2003})}\BibitemShut {NoStop}%
	\bibitem [{\citenamefont {Braggio}\ \emph {et~al.}(2006)\citenamefont
	  {Braggio}, \citenamefont {K\"onig},\ and\ \citenamefont
	  {Fazio}}]{BraggioPRL2006a}%
	  \BibitemOpen
	  \bibfield  {author} {\bibinfo {author} {\bibfnamefont {A.}~\bibnamefont
	  {Braggio}}, \bibinfo {author} {\bibfnamefont {J.}~\bibnamefont {K\"onig}}, \
	  and\ \bibinfo {author} {\bibfnamefont {R.}~\bibnamefont {Fazio}},\
	  }\href@noop {} {\bibfield  {journal} {\bibinfo  {journal} {Phys. Rev. Lett.}\
	  }\textbf {\bibinfo {volume} {96}},\ \bibinfo {pages} {026805} (\bibinfo
	  {year} {2006})}\BibitemShut {NoStop}%
	\bibitem [{\citenamefont {Flindt}\ \emph {et~al.}(2008)\citenamefont {Flindt},
	  \citenamefont {Novotn\'y}, \citenamefont {Braggio}, \citenamefont
	  {Sassetti},\ and\ \citenamefont {Jauho}}]{FlindtPRL2008a}%
	  \BibitemOpen
	  \bibfield  {author} {\bibinfo {author} {\bibfnamefont {C.}~\bibnamefont
	  {Flindt}}, \bibinfo {author} {\bibfnamefont {T.}~\bibnamefont {Novotn\'y}},
	  \bibinfo {author} {\bibfnamefont {A.}~\bibnamefont {Braggio}}, \bibinfo
	  {author} {\bibfnamefont {M.}~\bibnamefont {Sassetti}}, \ and\ \bibinfo
	  {author} {\bibfnamefont {A.-P.}\ \bibnamefont {Jauho}},\ }\href@noop {}
	  {\bibfield  {journal} {\bibinfo  {journal} {Phys. Rev. Lett.}\ }\textbf
	  {\bibinfo {volume} {100}},\ \bibinfo {pages} {150601} (\bibinfo {year}
	  {2008})}\BibitemShut {NoStop}%
	\bibitem [{\citenamefont {Flindt}\ \emph {et~al.}(2010)\citenamefont {Flindt},
	  \citenamefont {Novotn\'y}, \citenamefont {Braggio},\ and\ \citenamefont
	  {Jauho}}]{FlindtPRB2010a}%
	  \BibitemOpen
	  \bibfield  {author} {\bibinfo {author} {\bibfnamefont {C.}~\bibnamefont
	  {Flindt}}, \bibinfo {author} {\bibfnamefont {T.}~\bibnamefont {Novotn\'y}},
	  \bibinfo {author} {\bibfnamefont {A.}~\bibnamefont {Braggio}}, \ and\
	  \bibinfo {author} {\bibfnamefont {A.-P.}\ \bibnamefont {Jauho}},\ }\href@noop
	  {} {\bibfield  {journal} {\bibinfo  {journal} {Phys. Rev. B}\ }\textbf
	  {\bibinfo {volume} {82}},\ \bibinfo {pages} {155407} (\bibinfo {year}
	  {2010})}\BibitemShut {NoStop}%
	\bibitem [{\citenamefont {S\'anchez}\ and\ \citenamefont
	  {B\"uttiker}(2012)}]{SanchezEPL2012a}%
	  \BibitemOpen
	  \bibfield  {author} {\bibinfo {author} {\bibfnamefont {R.}~\bibnamefont
	  {S\'anchez}}\ and\ \bibinfo {author} {\bibfnamefont {M.}~\bibnamefont
	  {B\"uttiker}},\ }\href@noop {} {\bibfield  {journal} {\bibinfo  {journal}
	  {Europhys. Lett.}\ }\textbf {\bibinfo {volume} {100}},\ \bibinfo {pages}
	  {47008} (\bibinfo {year} {2012})}\BibitemShut {NoStop}%
	\bibitem [{\citenamefont {Gasparinetti}\ \emph {et~al.}(2014)\citenamefont
	  {Gasparinetti}, \citenamefont {Solinas}, \citenamefont {Braggio},\ and\
	  \citenamefont {Sassetti}}]{GasparinettiNJP2014a}%
	  \BibitemOpen
	  \bibfield  {author} {\bibinfo {author} {\bibfnamefont {S.}~\bibnamefont
	  {Gasparinetti}}, \bibinfo {author} {\bibfnamefont {P.}~\bibnamefont
	  {Solinas}}, \bibinfo {author} {\bibfnamefont {A.}~\bibnamefont {Braggio}}, \
	  and\ \bibinfo {author} {\bibfnamefont {M.}~\bibnamefont {Sassetti}},\
	  }\href@noop {} {\bibfield  {journal} {\bibinfo  {journal} {New J. Phys.}\
	  }\textbf {\bibinfo {volume} {16}},\ \bibinfo {pages} {115001} (\bibinfo
	  {year} {2014})}\BibitemShut {NoStop}%
	\bibitem [{Note2()}]{Note2}%
	  \BibitemOpen
	  \bibinfo {note} {In principle a full investigation at finite $ U_\alpha $ can
	  be performed, see Ref.~\protect \rev@citealpnum
	  {HusseinPRB2016a}.}\BibitemShut {Stop}%
	\bibitem [{Note3()}]{Note3}%
	  \BibitemOpen
	  \bibinfo {note} {The simplicity of this picture is due to the absence of the
	  double occupancy of the individual dots, which is energetically forbidden by
	  the strong interaction.}\BibitemShut {Stop}%
	\bibitem [{\citenamefont {Beenakker}(1991)}]{BeenakkerPRB1991a}%
	  \BibitemOpen
	  \bibfield  {author} {\bibinfo {author} {\bibfnamefont {C.~W.~J.}\
	  \bibnamefont {Beenakker}},\ }\href@noop {} {\bibfield  {journal} {\bibinfo
	  {journal} {Phys. Rev. B}\ }\textbf {\bibinfo {volume} {44}},\ \bibinfo
	  {pages} {1646} (\bibinfo {year} {1991})}\BibitemShut {NoStop}%
	\bibitem [{\citenamefont {Turek}\ and\ \citenamefont
	  {Matveev}(2002)}]{TurekPRB2002a}%
	  \BibitemOpen
	  \bibfield  {author} {\bibinfo {author} {\bibfnamefont {M.}~\bibnamefont
	  {Turek}}\ and\ \bibinfo {author} {\bibfnamefont {K.~A.}\ \bibnamefont
	  {Matveev}},\ }\href@noop {} {\bibfield  {journal} {\bibinfo  {journal} {Phys.
	  Rev. B}\ }\textbf {\bibinfo {volume} {65}},\ \bibinfo {pages} {115332}
	  (\bibinfo {year} {2002})}\BibitemShut {NoStop}%
	\bibitem [{\citenamefont {Nakpathomkun}\ \emph {et~al.}(2010)\citenamefont
	  {Nakpathomkun}, \citenamefont {Xu},\ and\ \citenamefont
	  {Linke}}]{NakpathomkunPRB2010a}%
	  \BibitemOpen
	  \bibfield  {author} {\bibinfo {author} {\bibfnamefont {N.}~\bibnamefont
	  {Nakpathomkun}}, \bibinfo {author} {\bibfnamefont {H.~Q.}\ \bibnamefont
	  {Xu}}, \ and\ \bibinfo {author} {\bibfnamefont {H.}~\bibnamefont {Linke}},\
	  }\href@noop {} {\bibfield  {journal} {\bibinfo  {journal} {Phys. Rev. B}\
	  }\textbf {\bibinfo {volume} {82}},\ \bibinfo {pages} {235428} (\bibinfo
	  {year} {2010})}\BibitemShut {NoStop}%
	\bibitem [{\citenamefont {Dubi}\ and\ \citenamefont
	  {Di~Ventra}(2011)}]{DubiRMP2011a}%
	  \BibitemOpen
	  \bibfield  {author} {\bibinfo {author} {\bibfnamefont {Y.}~\bibnamefont
	  {Dubi}}\ and\ \bibinfo {author} {\bibfnamefont {M.}~\bibnamefont
	  {Di~Ventra}},\ }\href@noop {} {\bibfield  {journal} {\bibinfo  {journal}
	  {Rev. Mod. Phys.}\ }\textbf {\bibinfo {volume} {83}},\ \bibinfo {pages} {131}
	  (\bibinfo {year} {2011})}\BibitemShut {NoStop}%
	\bibitem [{\citenamefont {Eltschka}\ \emph {et~al.}(2016)\citenamefont
	  {Eltschka}, \citenamefont {Thierschmann}, \citenamefont {Buhmann},\ and\
	  \citenamefont {Siewert}}]{EltschkaPSSA2016a}%
	  \BibitemOpen
	  \bibfield  {author} {\bibinfo {author} {\bibfnamefont {C.}~\bibnamefont
	  {Eltschka}}, \bibinfo {author} {\bibfnamefont {H.}~\bibnamefont
	  {Thierschmann}}, \bibinfo {author} {\bibfnamefont {H.}~\bibnamefont
	  {Buhmann}}, \ and\ \bibinfo {author} {\bibfnamefont {J.}~\bibnamefont
	  {Siewert}},\ }\href@noop {} {\bibfield  {journal} {\bibinfo  {journal} {Phys.
	  Status Solidi A}\ }\textbf {\bibinfo {volume} {213}},\ \bibinfo {pages} {626}
	  (\bibinfo {year} {2016})}\BibitemShut {NoStop}%
	\bibitem [{\citenamefont {Splettstoesser}\ \emph {et~al.}(2007)\citenamefont
	  {Splettstoesser}, \citenamefont {Governale}, \citenamefont {K\"onig},
	  \citenamefont {Taddei},\ and\ \citenamefont
	  {Fazio}}]{SplettstoesserPRB2007a}%
	  \BibitemOpen
	  \bibfield  {author} {\bibinfo {author} {\bibfnamefont {J.}~\bibnamefont
	  {Splettstoesser}}, \bibinfo {author} {\bibfnamefont {M.}~\bibnamefont
	  {Governale}}, \bibinfo {author} {\bibfnamefont {J.}~\bibnamefont {K\"onig}},
	  \bibinfo {author} {\bibfnamefont {F.}~\bibnamefont {Taddei}}, \ and\ \bibinfo
	  {author} {\bibfnamefont {R.}~\bibnamefont {Fazio}},\ }\href@noop {}
	  {\bibfield  {journal} {\bibinfo  {journal} {Phys. Rev. B}\ }\textbf {\bibinfo
	  {volume} {75}},\ \bibinfo {pages} {235302} (\bibinfo {year}
	  {2007})}\BibitemShut {NoStop}%
	\bibitem [{Note4()}]{Note4}%
	  \BibitemOpen
	  \bibinfo {note} {In order to have a finite nonlocal coupling $\Gamma _S$ both
	  the two local coupling $\Gamma _{S\alpha }$ with the superconductor need to
	  be finite as $\Gamma _S\propto \protect \sqrt {\Gamma _{SR}\Gamma
	  _{SL}}$.}\BibitemShut {Stop}%
	\bibitem [{\citenamefont {Rey}\ and\ \citenamefont {Sols}(2004)}]{ReyPRB2004a}%
	  \BibitemOpen
	  \bibfield  {author} {\bibinfo {author} {\bibfnamefont {M.}~\bibnamefont
	  {Rey}}\ and\ \bibinfo {author} {\bibfnamefont {F.}~\bibnamefont {Sols}},\
	  }\href@noop {} {\bibfield  {journal} {\bibinfo  {journal} {Phys. Rev. B}\
	  }\textbf {\bibinfo {volume} {70}},\ \bibinfo {pages} {125315} (\bibinfo
	  {year} {2004})}\BibitemShut {NoStop}%
	\bibitem [{\citenamefont {S\'anchez}\ \emph {et~al.}(2018)\citenamefont
	  {S\'anchez}, \citenamefont {Burset},\ and\ \citenamefont
	  {Yeyati}}]{SanchezPRB2018a}%
	  \BibitemOpen
	  \bibfield  {author} {\bibinfo {author} {\bibfnamefont {R.}~\bibnamefont
	  {S\'anchez}}, \bibinfo {author} {\bibfnamefont {P.}~\bibnamefont {Burset}}, \
	  and\ \bibinfo {author} {\bibfnamefont {A.~L.}\ \bibnamefont {Yeyati}},\
	  }\href@noop {} {\bibfield  {journal} {\bibinfo  {journal} {Phys. Rev. B}\
	  }\textbf {\bibinfo {volume} {98}},\ \bibinfo {pages} {241414} (\bibinfo
	  {year} {2018})}\BibitemShut {NoStop}%
	\bibitem [{\citenamefont {Kirsanov}\ \emph {et~al.}(2018)\citenamefont
	  {Kirsanov}, \citenamefont {Tan}, \citenamefont {Golubev}, \citenamefont
	  {Hakonen},\ and\ \citenamefont {Lesovik}}]{KirsanovArXiv2018a}%
	  \BibitemOpen
	  \bibfield  {author} {\bibinfo {author} {\bibfnamefont {N.~S.}\ \bibnamefont
	  {Kirsanov}}, \bibinfo {author} {\bibfnamefont {Z.~B.}\ \bibnamefont {Tan}},
	  \bibinfo {author} {\bibfnamefont {D.~S.}\ \bibnamefont {Golubev}}, \bibinfo
	  {author} {\bibfnamefont {P.~J.}\ \bibnamefont {Hakonen}}, \ and\ \bibinfo
	  {author} {\bibfnamefont {G.}~\bibnamefont {Lesovik}},\ }\href@noop {}
	  {\bibfield  {journal} {\bibinfo  {journal} {arxiv:1806.09838}\ } (\bibinfo
	  {year} {2018})}\BibitemShut {NoStop}%
	\bibitem [{\citenamefont {Pershoguba}\ and\ \citenamefont
	  {Glazman}(2019)}]{PershogubArXiv2019a}%
	  \BibitemOpen
	  \bibfield  {author} {\bibinfo {author} {\bibfnamefont {S.~S.}\ \bibnamefont
	  {Pershoguba}}\ and\ \bibinfo {author} {\bibfnamefont {L.~I.}\ \bibnamefont
	  {Glazman}},\ }\href@noop {} {\bibfield  {journal} {\bibinfo  {journal}
	  {arXiv:1901.10065}\ } (\bibinfo {year} {2019})}\BibitemShut {NoStop}%
	\end{thebibliography}
\end{document}